\documentclass[aps,prb,twocolumn,superscriptaddress,showpacs,amsmath,amssymb, amsfonts,10pt,longbibliography]{revtex4-2}

\usepackage{amsmath}
\usepackage{amsfonts}
\usepackage{amssymb}
\usepackage{graphicx}
\usepackage{color}
\usepackage[dvipsnames]{xcolor}
\usepackage{dsfont}
\usepackage[caption=false]{subfig}
\usepackage{bm}
\usepackage{bbold}
\usepackage{fancyvrb}

\usepackage{mathtools}

\usepackage{xcite} % bibliography

\usepackage[normalem]{ulem}
\usepackage[colorlinks,bookmarks=false,urlcolor=blue, citecolor=blue, linkcolor = blue]{hyperref}

\usepackage{pgfplots} % main
\usepackage{pgfmath} % for some math functions
\pgfplotsset{width=0.7\textwidth,compat=1.9} % default width and version
\usepgfplotslibrary{fillbetween} % used once, fill area between 2 lines
\usepgfplotslibrary{external} % save fig external, not compile every time
\pgfkeys{/pgf/number format/.cd,1000 sep={\,}} % format of numbers, maybe not needed
\usetikzlibrary{pgfplots.groupplots} % axes breaks
\usetikzlibrary{shapes,arrows}

\newcommand{\dd}{\mathrm{d}}
\newcommand{\ee}{\mathrm{e}}
\newcommand{\ii}{\mathrm{i}}

\begin{document}

\title{Quantum Wires with Local Particle Loss: Transport Manifestations of Fluctuation-Induced Effects}

\author{Marcel Gievers}
\email{m.gievers@lmu.de}
\affiliation{Arnold Sommerfeld Center for Theoretical Physics, Center for NanoScience, and Munich Center for Quantum Science and Technology, Ludwig-Maximilians-Universität München, 80333 Munich, Germany}
\affiliation{Max-Planck-Institute  of  Quantum  Optics,  85748  Garching,  Germany}

\author{Thomas Müller}
\affiliation{Institute for Theoretical Physics, University of Cologne, 50937 Cologne, Germany}

\author{Heinrich Fröml}
\affiliation{Institute for Theoretical Physics, University of Cologne, 50937 Cologne, Germany}

\author{Sebastian Diehl}
\affiliation{Institute for Theoretical Physics, University of Cologne, 50937 Cologne, Germany}

\author{Alessio Chiocchetta}
\affiliation{Institute for Theoretical Physics, University of Cologne, 50937 Cologne, Germany}

\begin{abstract}
We investigate the transport properties of a quantum wire of weakly interacting fermions in the presence of local particle loss. We calculate current and conductance in this system due to applied external chemical potential bias that can be measured in experimental realizations of ultracold fermions in quasi one-dimensional traps. Using a Keldysh field theory approach based on the Lindblad equation, we establish a perturbative scheme to study the effect of imbalanced reservoirs. Logarithmically divergent terms are resummed using a renormalization group method, and a novel powerlaw behavior for the conductance as a function of the potential bias across the wire is found. In contrast to the equilibrium case of a potential barrier in a Luttinger liquid, the conductance exhibits a scaling behavior, which depends on the interaction strength and on the loss probability. Repulsive interactions reduce the conductance of the wire while attractive interactions enhance it. However, perfect reflectivity and transparency are only reached in the absence of particle loss. 
\end{abstract}

\date{\today}

\maketitle

\section{Introduction}
\label{sec:introduction}

One-dimensional systems provide among the most intriguing manifestations of quantum manybody effects. The interplay between gapless quantum fluctuations and low-dimensional geometry gives rise to unexpected collective behaviors: for instance, the quasiparticles are bosonic in nature, irrespectively of the underlying particle statistics, a separation of spin and charge degrees of freedom occurs, and observables exhibit scaling behavior controlled by power laws~\cite{giamarchi2003quantum,gogolin2004bosonization}. This universal behavior of such Luttinger liquids~\cite{giamarchi2003quantum,gogolin2004bosonization} was experimentally observed in a number of platforms, including nanotubes~\cite{Bockrath99,Yao99,Cao05}, quantum Hall edges~\cite{Chang03}, cold atoms~\cite{Yang17,Esslinger18,Yang18}, quantum circuits~\cite{Anthore18}, antiferromagnetic spin chains ~\cite{Lake2005}, and spin-ladder systems~\cite{Dagotto1999,Giamarchi2008}. \par
By virtue of the one-dimensional geometry, the most natural experimental protocol to probe these systems is via transport measurements. This is achieved by coupling the ends of the wire to two reservoirs (``leads''), imbalanced in terms of chemical potentials or temperature~\cite{Nazarov2009} while in ultracold-atomic wires this imbalance is created by directly changing, for instance, density or temperature in the reservoirs~\cite{Brantut2012,Krinner2015,krinner2017two,lebrat2019local,Husmann2018,Krinner8144}. More recently, ultracold atomic experiments were performed measuring the current through a superconducting quantum point contact~\cite{Visuri2023,Huang2023,fabritius2024irreversible}. In particular, these experiments typically operate outside the regime of linear transport such that linear response is not sufficient to describe the observed current-bias characteristics and an explicit nonequilibrium manybody theory is required. \par
One of the most striking effects visible in transport experiments concerns the presence of impurities within the wire. Seminal work by Kane and Fisher~\cite{kane1992transmission,kane1992transport,Kane1992rapid} indeed demonstrated that the conductance of a quantum wire is dramatically modified by the presence of an impurity. In fact, gapless excitations can enhance or suppress the backscattering due to the impurity, leading to dramatic signatures in the wire's conductance. For instance, a small impurity leads to complete suppression of the conductance through the wire, thus dramatically modifying the Landauer paradigm. Furthermore, the temperature and voltage dependence of the conductance follows power laws due to the gapless nature of a Luttinger liquid.  \par 
Quantum wires and quantum dots coupled to external reservoirs have been intensively studied within the last decades~\cite{LeHur2005Unification,Zhang2021Nonequilibrium}. Experimentally these systems are typically realized in carbon nanotubes, where the dissipation of energy plays a crucial role~\cite{Bomze2009Resonant,mebrahtu2012quantum,mebrahtu2013observation,Liu2014Tunable}.
In recent years, a series of experiments in cold atomic systems sparked new interest in the role of a new type of impurities in one-dimensional geometries. There, highly energetic and spatially narrow beams were shone on the gases, creating a localized single-particle loss~\cite{barontini2013controlling,labouvie2016bistability,mullers2018coherent,lebrat2019local,corman2019quantized}.\par
These resulting lossy impurities drive the system out of thermal equilibrium and unveil novel physical effects, such as the quantum Zeno effect (QZE), observed in a Bose gas~\cite{barontini2013controlling,labouvie2016bistability} and theoretically described in several works~\cite{Brazhnyi2009,Shchesnovich2010May,Shchesnovich2010Oct,Witthaut2011,Barmettler2011,Zezyulin2012,Barmettler2011,Kepesidis2012,Kordas2013,Kiefer2017,Kunimi2019,Bychek2019}. The interplay between lossy impurities and coherent dynamics was shown to lead to additional phenomena qualitatively different from their equilibrium counterparts, or with no counterpart at all, e.g., fluctuation-induced quantum Zeno effect~\cite{Froeml2019,Froeml2020,muller2021shape}, dynamical phase transitions~\cite{SELS2020168021,buca2020dissipative,Ueda2021,Haque2020,Saleur2020}, orthogonality catastrophe~\cite{Federico1,Berdanier}, and exotic nonequilibrium steady states~\cite{Schnell2017,Lapp2019,Yanay2018,Yanay2020, Krapivsky_2019,Dutta1,Krapivsky_2020,Dutta2,alba2022noninteracting,alba2022unbounded}. Further promising directions were investigated for mobile~\cite{Piazza2021} and dephasing impurities~\cite{Dolgirev,federico2}. \par
While interaction effects in an unbiased system~\cite{Froeml2019,Froeml2020,muller2021shape} and nonlinear transport properties in noninteracting lattice systems~\cite{Uchino2022Comparative,VisuriPRB2023,VisuriPRL2023} have been studied, a study of the interplay of local particle loss, interactions and an explicit external voltage bias has only been conducted in the context of superconducting quantum point contacts~\cite{Visuri2023,Huang2023,fabritius2024irreversible}. The aim of our work is to provide a theory for the transport in dissipative and interacting quantum wires by generalizing a renormalization group approach for the treatment of junctions of interacting quantum wires~\cite{Aristov_2014,Aristov_2017,Aristov_2018,Aristov_2019,nosov2020tunneling} to the dissipative case. Note that in the present work, \emph{dissipation} actually refers to the loss of particles. \par 
We investigate the transport properties of a fermionic quantum wire with a local dissipative impurity coupled to two imbalanced reservoirs (cf.\ Fig.~\ref{fig:Dissipative-wire-model}). To include nonlinear effects and adequately describe the particle loss of the system, we apply the Keldysh formalism solving the Lindblad equation~\cite{sieberer2016keldysh,sieberer2023universality}. In this framework, we are able to compute physical observables in the steady state determined by the interplay between the reservoir imbalance and the lossy impurity. In particular, this allows us to evaluate the currents in the system, and perturbative corrections thereof, due to the interparticle interaction. \par
In previous works~\cite{Froeml2019,Froeml2020,muller2021shape}, the renormalized scattering coefficients of a lossy impurity were calculated at the Fermi surface. There, fully universal fluctuation-induced phenomena were retrieved. In this work, on the contrary, we directly compute physical observables, i.e., the current and conductance of the wire at imbalanced reservoirs. We show that, in presence of local particle loss, modes below the Fermi energy contribute to the transport enabled by the replenishing of holes deep in the Fermi sea created by the lossy impurity (cf.\ Fig.~\ref{fig:transport-processes}), which leads to a modified behavior.\par
Our main result is a renormalization group (RG) analysis of the transport properties in the presence of the dissipative impurity. The RG scaling parameter is given in the limit of small voltages. We find that a small impurity does not completely suppress the conductance, in contrast to the equilibrium case. Instead, the conductance is renormalized to a value dependent on the microscopic losses. Remarkably, this renormalized conductance exhibits a nonmonotonic behavior as a function of the impurity strength, realizing a novel manybody incarnation of the quantum Zeno effect. \par
Furthermore, we derive an algebraic behavior of the conductance in terms of the voltage. As the RG flow of the conductance depends on more microscopic parameters, also the powerlaw exponent depends on the details of the lossy impurity. This is qualitatively different from the equilibrium counterpart for transport through a potential barrier, where the scaling form is independent of the microscopic details of the impurity and is solely determined by the interactions in the lead. \par
The modified behavior of the conductance in a dissipative wire is evoked by the interplay of incoherent particle loss and particle interactions and therefore provides a way to experimentally characterize the full physics of lossy impurities by conductance measurements. \par
This article is structured as follows: In Sec.~\ref{sec:Model} we present our model of a wire with local particle loss connected to two Fermi-liquid reservoirs. In Sec.~\ref{sec:Non-interacting} we discuss the transport properties in the noninteracting case. In Sec.~\ref{sec:First-order} we then derive corrections for the currents to first order in particle-particle interactions. Higher-order effects are taken into account by the renormalization group analysis of the conductance, which we discuss in Sec.~\ref{sec:Renormalization-Conductances}. \par

\section{The Model \label{sec:Model}}

\begin{figure}[t!]
	\centering
	\includegraphics[width=0.48\textwidth]{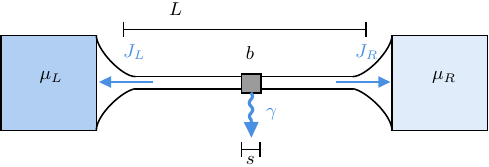}
	\caption{Illustration of a quantum wire coupled to Fermi liquid reservoirs with chemical potentials $\mu_L$ and $\mu_R$. The impurity consists of a coherent barrier of strength $b$ and a dissipation strength $\gamma$, which controls the local particle loss. The extension of the impurity $s$ is much smaller than the length $L$ of the interacting wire. The two currents $J_L$ and $J_R$ are conventionally defined to be positive when pointing out of the wire. An imbalance $V = \mu_L - \mu_R > 0$ generates a net current $J_\mathrm{net} = (J_R-J_L)/2$ flowing from the left to the right.}
	\label{fig:Dissipative-wire-model}
\end{figure}

We consider a one-dimensional gas of spinless fermions with mass $m$, interacting via a short-ranged interaction $g(x-y)$ over a region $|x| < L/2$. The system is described by the Hamiltonian 
\begin{equation}
\hat{H}_\text{wire} = \int_x \hat{\psi}^\dagger(x)\frac{-\partial_x^2}{2m}\hat{\psi}(x) + \frac{1}{2}\int_{x,y}\!\!\!g(x-y)\hat{n}(x)\hat{n}(y),
\label{eq:Hamiltonian}
\end{equation}
with $\int_x \equiv \int_{-L/2}^{L/2} \dd x$, $\hat{n} =\hat{\psi}^\dagger\hat{\psi}$ the density operator, and $\hat{\psi}^{\dagger}, \hat{\psi}$ the fermionic creation and annihilation operators. Throughout this paper, we set the natural constants $e$ and $\hbar$ to unity. The boundaries of the wire are assumed to be adiabatically connected to particle reservoirs with chemical potentials, $\mu_L$ and $\mu_R$ causing a chemical potential imbalance $V=\mu_L-\mu_R$ (also referred to as voltage), but equal temperature $T$. Under this assumption, no scattering takes place at the interfaces between the wire and the reservoirs: while this may not be justified for solid-state wires~\cite{Maslov1995,Ponomarenko1995,Safi1995}, it captures the geometry of ultracold quantum wires~\cite{Husmann2018,lebrat2019local,Krinner2015}   %\marcel{later discussion $L\to 2L$, $s\to 2s$}
The reservoirs determine the particle density and momentum distribution in the wire, as discussed in further detail in Sec.~\ref{sec:Non-interacting}. In general, different reservoirs may give rise to peculiar phenomena, such as Andreev reflection for superfluid reservoirs~\cite{Safi1995}, or anomalous transport for reservoirs of unitary Fermi gases~\cite{Husmann2018}.  \par
In the following, we will consider ideal Fermi gases as reservoirs, for the sake of simplicity. For vanishing interactions ($g=0$) the particles in the wire with a positive momentum $k>0$ (respectively, negative momentum $k<0$), originating from the left (respectively, right) reservoir, are populated according to the Fermi distributions $f_L(k)$ (respectively, $f_R(k)$), with
$f_i(k) = 1/\left(1+\exp[(\varepsilon_k-\mu_i)/T]\right)$, $i = L,R$, and the quadratic dispersion relation $\varepsilon_k = k^2/(2m)$ of free particles.
The coherent and dissipative impurities are centered at $x=0$, and have arbitrary spatial profiles vanishing for $|x|>s/2$, with $s\ll L$. The coherent part of the impurity is modeled by the Hamiltonian
\begin{equation}
	\hat{H}_\mathrm{imp} = \int_x   b(x)\hat{\psi}^\dagger(x)\hat{\psi}(x),
\end{equation}
with $b(x)$ a function describing its spatial profile. The dissipative part of the impurity is, instead, described by Markovian single-particle loss. To model this, it is convenient to describe the dynamics of the system by a quantum master equation~\cite{breuer_theory_2007} for the system's density matrix $\hat\rho$ as
\begin{equation}
\label{eq:Lindblad-equation}
\partial_t {\hat{\rho}} = -\mathrm{i} [\hat{H}, \hat{\rho} ] + \mathcal{D}[ \hat{\rho}],
\end{equation}
with the total Hamiltonian $\hat{H}=\hat{H}_{\text{wire}}+\hat{H}_\mathrm{imp}$ and the dissipator
\begin{equation}
\label{eq:Lindblad}
	\mathcal{D} [\hat{\rho} ]=  \int_x \gamma(x) \left[ \hat{\psi}(x) \hat{\rho} \hat{\psi}^\dagger(x) - \frac{1}{2}\left\{ \hat{\psi}^\dagger(x) \hat{\psi}(x), \hat{\rho} \right\} \right],
\end{equation}
The full model is illustrated by Fig.~\ref{fig:Dissipative-wire-model}.
In the rest of this work, we will assume the system to be in the stationary state determined by the balance between particles refilled and absorbed by the reservoirs, and particles lost at the dissipative impurity. 

\section{Dissipative transport in the noninteracting case \label{sec:Non-interacting}}

In the absence of interactions ($g=0$), the model is exactly solvable. In this section, we provide an exact expression for certain observables relevant to experimental platforms.

\subsection{Multiple-wire basis and Green's functions}

To evaluate observables, it is convenient to use a different basis for the fermionic operators $\psi^\dagger(x), \psi(x)$:
\begin{equation}
\hat{\psi}(x) = 
\begin{cases}
\hat{\psi}_L(-x) & \text{for } x<0\\
\hat{\psi}_R(x) & \text{for } x>0
\end{cases}, 
\label{eq:LR-basis}
\end{equation}
where $\hat{\psi}^\dagger_i, \hat{\psi}_i$ with $i = L,R$ denote the creation and annihilation operators for particles being at the left (respectively, right) side of the impurity. Note that the arguments $x$ of the operators $\hat{\psi}_i$ are always positive, as they measure the distance from the impurity. \par
A central role is played by the retarded, advanced, and Keldysh Green's function, defined, respectively, as~\cite{kamenev_2011} 
\begin{subequations}
\label{eq:Green-definition}
\begin{align}
\mathcal{G}^\mathrm{R}(1,2) &=-\mathrm{i}\theta(t_1-t_2)\left\langle\left\{\hat{\psi}(1),\hat{\psi}^\dagger(2)\right\}\right\rangle,\\
\mathcal{G}^\mathrm{A}(1,2)
&=\phantom{-}\mathrm{i}\theta(t_2-t_1)\left\langle\left\{\hat{\psi}(1),\hat{\psi}^\dagger(2)\right\}\right\rangle,\\
\mathcal{G}^\mathrm{K}(1,2) &=-\mathrm{i}\left\langle\left[\hat{\psi}(1),\hat{\psi}^\dagger(2)\right]\right\rangle.
\end{align}
\end{subequations}
Here, $\hat\psi(\alpha)$ denotes a fermionic operator evaluated at a general argument $\alpha$ with corresponding time $t_\alpha$. In particular, $\mathcal{G}^\mathrm{R}$ and $\mathcal{G}^\mathrm{A}$ describe the single-particle response of the system to an external perturbation while $\mathcal{G}^\mathrm{K}$ represents correlations. At equal times, up to the imaginary prefactor, $\mathcal{G}^\mathrm{K}$ describes the single-particle covariance matrix, whose eigenvalues are the particle occupation numbers~\cite{sieberer2016keldysh,sieberer2023universality}. Thus many physical observables can be evaluated from the Green's functions (e.g., densities and currents as discussed below). In the stationary state, the Green's functions in frequency and real space can be computed by solving the corresponding Dyson equations (see Apps.~\ref{sec:Regularization-Green} and~\ref{sec:Dyson}). The closed form of the resulting Green's functions in frequency-position representation, i.e., $\mathcal{G}^{\mathrm{R,A,K}}_\omega(i,x|j,y)$ with $i,j = L,R$, is cumbersome and reported in App.~\ref{sec:Dyson}, Eq.~\eqref{eq:Impurity-Green}. However, a remarkable feature of the Green's functions is their dependence on the scattering matrix $S$, which encodes the scattering properties of the impurity as
\begin{equation}
S(k) = 
\begin{pmatrix}
r_k^L & t_k \\ 
t_k & r_k^R
\end{pmatrix},
\label{eq:Scattering-matrix}
\end{equation}
with $r_k^L$ (respectively, $r_k^R$) the reflection amplitudes from the left (respectively, right) side of the impurity and $t_k$ the transmission coefficient. \par
The values of the coefficients $r_k^L$, $r_k^R$, and $t_k$ can be evaluated by solving a non-Hermitian Schr\"{o}dinger equation associated with the impurity~\cite{Froeml2019,Froeml2020}. While $r_k^L = r_k^R$ for inversion-symmetric scenarios (i.e., invariant under $x\to -x$), $t_k$ is always direction-independent~\cite{Valentin-bachelor}. Here, solving a non-Hermitian Schrödinger equation to derive scattering properties is only justified in the absence of interactions. For lossy impurities, the scattering matrix $S(k)$, Eq.~\eqref{eq:Scattering-matrix}, is not unitary anymore. \par
We treat symmetric impurities, i.e., $r_k^L=r_k^R$, however, as outlined below, the interplay of interactions $g$, dissipation $\gamma$ and a finite voltage $V$ renormalizes a symmetric impurity to an asymmetric one, $r_k^L\neq r_k^R$ as the inversion-symmetry is broken by the voltage. As an example for the scattering amplitudes, consider a pointlike impurity, i.e., $b(x)= b\delta(x)$ and $\gamma(x)=\gamma\delta(x)$, which will also be used later (cf.\ Sec.~\ref{sec:Renormalization-Conductances}). In this case, the coefficients of the scattering matrix $S$ are given by
\begin{equation}
		r_k = \frac{-b+\mathrm{i}\gamma/2}{b-\mathrm{i}\gamma/2-\mathrm{i}v_k}, \quad
		t_k = \frac{-\mathrm{i}v_k}{b-\mathrm{i}\gamma/2-\mathrm{i}v_k}.
	\label{eq:Scattering-delta}
\end{equation}
The dissipative behavior of the impurity is best recognized by introducing the loss probability $\eta_k$
\begin{equation}
	\eta_k = 1-|r_k|^2-|t_k|^2 = \frac{v_k\gamma}{b^2+(\gamma/2 +v_k)^2}.
	\label{eq:eta_k-condition}
\end{equation}
The loss probability $\eta_k$ depends nonmonotonically on the dissipation strengths $\gamma$, which is a single-particle manifestation of the QZE~\cite{Froeml2019,Froeml2020}. \par

\subsection{Density profile and Friedel oscillations}
\label{sec:density}

The density profile in the quantum wire is an experimentally accessible quantity bearing signatures of the underlying scattering processes. In fact, the presence of an impurity leads to the emergence of Friedel oscillations, which, through particle interactions, modify crucially the transmission properties of the impurity~\cite{matveev1993tunneling,Froeml2019}. \par
\begin{figure}[t!]
	\centering
	\includegraphics[width=0.45\textwidth]{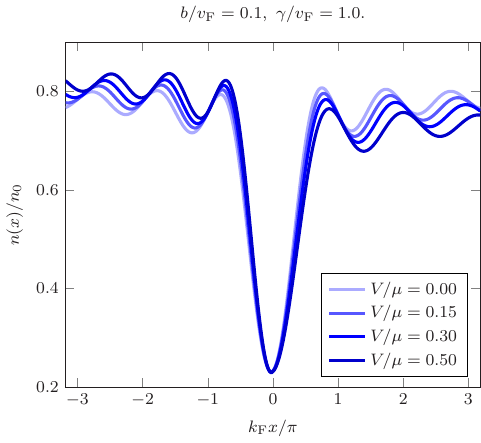}
	\caption{Friedel oscillations for a finite coherent strength $b$, dissipation strength $\gamma$ and different voltages $V$ at zero temperature. As reference scale for the density profiles $n(x)$ we take the background density for the coherent case $n_0 \equiv {n}^\mathrm{back}(V,b,\gamma=0)$.}
	\label{fig:Friedel-oscillations}
\end{figure}

The density profile $n_i(x)=\langle\hat{\psi}_i^\dagger(x)\hat{\psi}_i(x)\rangle$ on either side of the impurity $i=L,R$ can be evaluated directly from the Keldysh Green's function [cf.\ Eq.~\eqref{eq:Density-definition}]. It is convenient to decompose it as $n_i(x)={n}^\mathrm{back}_i + {n}^\mathrm{osc}_i(x)$, with ${n}^\mathrm{back}_i$ the average constant density, and ${n}^\mathrm{osc}_i(x)$ the spatially oscillating parts. These read
\begin{subequations}
\label{eq:densities}
\begin{align}
{n}^\mathrm{back}_{i}  &=\int_{k>0} \bigg[ \left(|r^{i}_k|^2+1\right)f_{i} + |t_k|^2 f_{\bar i} \bigg], \label{eq:background-density}\\
{n}^\mathrm{osc}_{i}  &=  2 \int_{k>0} \mathrm{Re}[r^{i}_k \ee^{2\mathrm{i} k x}] f_{i},
\label{eq:Friedel}
\end{align}
\end{subequations}
where $\bar{i}$ denotes the contrary reservoir index (i.e., $\bar L = R$ and $\bar R = L$). For $k_i\,x \gg 1$, where $k_i=\sqrt{2m\mu_i}$ is the Fermi momentum of the left (respectively, right) reservoir, the oscillating contribution in Eq.~\eqref{eq:Friedel} acquires the simple form ${n}^\mathrm{osc}_i(x) \propto  \sin(2 k_i x )/x$ (cf.\ Fig.~\ref{fig:Friedel-oscillations}). This form of Friedel oscillations relies on the sharp Fermi edge and a smooth momentum dependence of the reflection coefficient $r_k^{L,R}$. Remarkably, due to the different Fermi momenta in the reservoirs, the oscillation period, given by $\pi/k_i$, is different on the two sides of the impurities. Thus, a finite voltage induces an asymmetry in the density profile. In absence of impurities, i.e., $b = 0 = \gamma$, the Friedel oscillations vanish and the densities ${n}^\mathrm{back}_{L}$ and ${n}^\mathrm{back}_{R}$ are given by the average of the reservoir populations. Conversely, ${n}^\mathrm{back}_{L} \neq {n}^\mathrm{back}_{R}$ in presence of an impurity, as shown in Fig.~\ref{fig:Friedel-oscillations}, and a finite density imbalance is established in the two wire sections. \par 
The background density ${n}^\mathrm{back}_i$ exhibits a nonmonotonic dependence on the dissipation strength $\gamma$, revealing the quantum Zeno effect, as, for large  $\gamma$, the background density ${n}^\mathrm{back}_{L,R}$ rises again towards the value of the coherent system~\cite{Froeml2019,Froeml2020}. \par

Finally, we notice that Eq.~\eqref{eq:densities}, for $V=0$, acquires exactly the same form which was derived in the quasistationary state for a quench of a localized loss in an isolated quantum wire~\cite{Froeml2019,Froeml2020}. This confirms the picture that, in that case, the far ends of the wire act as reservoirs of particles, balancing the losses induced by the dissipative impurity. \par

\subsection{Currents and the quantum Zeno effect}
\label{sec:currents}

\begin{figure}
	\includegraphics[width=0.48\textwidth]{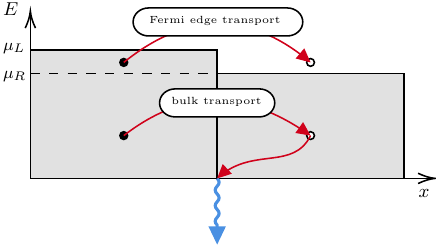}
	\caption{Processes contributing to transport through a lossy impurity. In addition to the particles at the Fermi edge, described by the standard Landauer-Büttiker formula, the bulk fermions contribute following holes in the Fermi sea that emerge due to the loss of fermions at all momenta. In the absence of particle loss, these processes are not possible due to Pauli blocking in a completely filled Fermi sea.}
\label{fig:transport-processes}
\end{figure}

The average current flowing from the wire toward the reservoir $i=L,R$ (cf.\ Fig.~\ref{fig:Dissipative-wire-model}) is given by $J_i(x) =\mathrm{Im}\langle\hat{\psi}_i^\dagger(x)\partial_x\hat{\psi}_i(x)\rangle/m$. Following the conventions given in Eq.~\eqref{eq:LR-basis}, the expression for the current $J_i(x)$ is positive when it is pointing toward the reservoir $i$ and negative when it is pointing toward the impurity. It can be evaluated directly from the equal-time Keldysh Green's function [cf.\ Eq.~\eqref{eq:Current-definition}] and yields
\begin{equation}
	J_i = \int_{k>0} v_k \sum_j \left(|S_{ij}|^2-\delta_{ij}\right)f_j.
	\label{eq:Current-j}
\end{equation}
Equation~\eqref{eq:Current-j} is analogous to the Landauer-B\"{u}ttiker formula~\cite{Nazarov2009}. Note that $J_i$ does not depend on the position $x$, but just on the side of the impurity. For an inversion-symmetric impurity, the currents acquire the simpler form
\begin{equation}
		J_i = \int_{k>0} v_k \left[|t_k|^2(f_{\bar{i}}(k)-f_i(k))-\eta_k f_i(k)\right],
	\label{eq:non-interacting:J_1-J_2}
\end{equation}
which shows that currents are induced not only by the imbalance between the reservoirs' chemical potentials but also from losses induced by the impurity, encoded by $\eta_k$ [cf.\ Eq.~\eqref{eq:eta_k-condition}]. In particular, the first term in the integral is mainly determined by states near the Fermi energy, as in the usual Landauer-B\"{u}ttiker theory. Conversely, every state below the Fermi energy contributes to the second term in the integral, which is of purely dissipative nature, as noted in Ref.~\onlinecite{corman2019quantized}. This additional term is relevant for the net transport through the impurity since it is asymmetric in the presence of a potential bias. This may be understood in terms of the mechanism illustrated in Fig.~\ref{fig:transport-processes}: If particles are lost from the side at lower potential, there emerges a hole in the Fermi sea that can be filled by a particle tunneling through the impurity, therefore leading to a net flow of particles which adds to the transport of particles at the Fermi edge. 
It is convenient to introduce the loss current as
\begin{equation}
J_\mathrm{loss} = -J_L-J_R = \int_{k>0} v_k \eta_k (f_L+f_R),
\label{eq:non-interacting:J_loss}
\end{equation}
which quantifies the rate of particles lost from the wire due to the lossy impurity. \par
\begin{figure}[t!]
	\centering
	\includegraphics[width=0.45\textwidth]{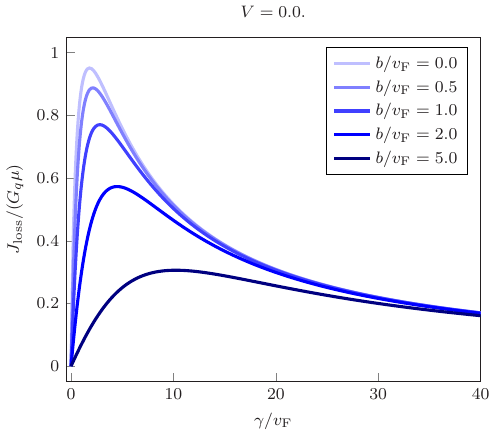}
	\caption{
	Loss current $J_\mathrm{loss}$ as function of the dissipation strength $\gamma$ for different coherent strengths $b$. For large $\gamma$ the extinction of the loss current happens due to the single-particle quantum Zeno effect. A coherent impurity strength does not qualitatively alter the non-monotonous behavior of the loss current through the dissipative impurity.
	}
	\label{fig:Single-particle-QZE}
\end{figure}
It is evident in Eq.~\eqref{eq:non-interacting:J_loss} that the value of the loss current takes into account all the momenta, not only those close to the Fermi surface. As it includes the loss probability $\eta_k$ as the only scattering quantity, the loss current $J_\mathrm{loss}$ exhibits a nonmonotonic dependence on the dissipation strength $\gamma$. In particular, for large $\gamma$, the loss current $J_\mathrm{loss}$ is suppressed which is understood by means of the single-particle quantum Zeno effect. This effect was experimentally verified in Bose gases~\cite{barontini2013controlling,labouvie2016bistability}. The typical nonmonotonic behavior as a function of $\gamma$ is displayed in Fig.~\ref{fig:Single-particle-QZE}. Recently, such a behavior has also been found in transport through monitored quantum dots~\cite{FerreiraTransport2024}. \par
Let us now introduce the central physical quantity of our work, namely the differential conductance, which gives a measure of how the current reacts to a change of the reservoirs' imbalance $V = \mu_L-\mu_R$:
\begin{equation}
\label{eq:diff_conductance}
G_{i}=\frac{\partial J_{i}}{\partial V},
\end{equation}
At low temperatures, the Fermi edge is sharp, i.e., $f_{i}(k)~\approx~\theta(\mu_{i}-\varepsilon_k)$, so the derivative of the current $J_i$, Eq.~\eqref{eq:non-interacting:J_1-J_2}, with respect to $V$ peaks up the scattering probabilities at the respective Fermi momenta, since $\partial f_i/\partial V \approx \mp \delta(k-k_i)/(2v_k)$. Thus, the differentiated conductance acquires the simple form
\begin{equation}
   G_{i} \approx \pm \frac{1}{2} \left(|t_{\bar{i}}|^2+|t_{i}|^2+\eta_{i}\right),
   \label{eq:G_L,R(t_L,R}
\end{equation}
where the scattering coefficients $t_k$ and $\eta_k$ are taken at the Fermi momentum $k_i$ of the corresponding reservoirs, i.e., $t_i \equiv t_{k_i}$ and $\eta_i \equiv \eta_{k_i}$.
Eq.~\eqref{eq:G_L,R(t_L,R} is given in units of the conductance quantum $G_q=e^2/h$. From now on, we always give conductances in units of $G_q$ omitting a factor of $2\pi$. If we further assume that the microscopic scattering coefficients depend smoothly on momentum, then for small voltages, $V=\mu_L-\mu_R\ll(\mu_L+\mu_R)/2\equiv\mu$, we can omit the difference between $k_L$ and $k_R$ in the argument of the scattering amplitudes and evaluate them at the average Fermi momentum $k_\mathrm{F}=\left(k_L+k_R\right)/2$ with $t \equiv t_{k_\mathrm{F}}$ and $\eta \equiv \eta_{k_\mathrm{F}}$. Then there is only one independent conductance, namely $G = |t|^2+\eta/2 = G_L = - G_R$. Let us emphasize that eventually we are interested in the regime of small voltages, $V\to 0$, as there the RG picture is valid. \par
The states from the bulk contribute to the individual currents $J_L$ and $J_R$ [cf.\ Fig.~\ref{fig:transport-processes} and Eq.~\eqref{eq:non-interacting:J_1-J_2}], however these contributions exactly cancel each other in the net current through the wire, $J_\mathrm{net} = {(J_R-J_L)/2}=\int_{k>0} v_k (|t_k|^2+\eta_k/2)(f_L-f_R)$. This is the reason why the corresponding conductance $G$, Eq.~\eqref{eq:G_L,R(t_L,R}, depends on the scattering probabilities at the Fermi level as this is sensitive to a change of chemical potentials. We will argue below that in presence of interactions, Eq.~\eqref{eq:G_L,R(t_L,R} is not applicable anymore. In other words, interactions renormalize the conductance $G$ in a different way than the scattering probabilities $|t_i|^2$ and $\eta_i$ when particle loss is included. \par

\section{First-order perturbative correction to currents} 
\label{sec:First-order}

In this section, we outline the perturbation theory treating small interactions $g$ in the nonequilibrium steady state, using the Keldysh formalism~\cite{kamenev_2011}. This constitutes one of the main results of our article, as it provides a systematic framework to approach transport in a quantum wires with a lossy impurity. Since in the dissipative case the currents are also carried by states far below the Fermi energy (cf.\ Fig.~\ref{fig:transport-processes}), we do not rely on a linearized theory~\cite{Aristov_2014,Aristov_2017,Aristov_2018,Aristov_2019,nosov2020tunneling}, but include the whole quadratic dispersion relation. The results from first-order perturbation theory provide the foundation for the RG analysis in Sec.~\ref{sec:Renormalization-Conductances}. Readers who are not interested in technical details may directly proceed with that section. \par

\subsection{Perturbation theory in Keldysh formalism}

In Keldysh formalism, one makes use of the fact that the density matrix $\hat\rho(t)$, which in our case is given by the quantum master equation~\eqref{eq:Lindblad-equation}, evolves along two time strings. These are conveniently labeled as forward $+$ and backward branch $-$. After applying a Trotter decomposition of the partition function $\mathcal{Z} = \mathrm{tr}\left[\hat{\rho}(t)\right]$, naturally two distinct fields $\psi^+$ and $\psi^-$ evolve from the forward and backward time contours~\cite{sieberer2016keldysh,sieberer2023universality}. In the corresponding Keldysh field theory, the expressions for the three Green's functions~\footnote{$i=R,L$ is the respective wire index which can be generalized to an arbitrary number of wires meeting in a junction.} are calculated by the classical and quantum fields $\psi^c = \left(\psi^++\psi^-\right)/\sqrt{2}$ and $\psi^q=\left(\psi^+-\psi^-\right)/\sqrt{2}$ from the expectation values~\cite{kamenev_2011}
\begin{subequations}
    \label{eq:Keldysh-field-theory_Green's-functions}
	\begin{align}
		\mathcal{G}^\mathrm{R}_\omega(i,x|j,y) &= -\mathrm{i}\left\langle\psi_i^c(\omega,x)\bar{\psi}_j^q(\omega,y)\right\rangle,\\
		\mathcal{G}^\mathrm{A}_\omega(i,x|j,y) &= -\mathrm{i}\left\langle\psi_i^q(\omega,x)\bar{\psi}_j^c(\omega,y)\right\rangle,\\
		\mathcal{G}^\mathrm{K}_\omega(i,x|j,y) &= -\mathrm{i}\left\langle\psi_i^c(\omega,x)\bar{\psi}_j^c(\omega,y)\right\rangle .
	\end{align}
\end{subequations}
\begin{figure}[t!]
	\centering
	\includegraphics[width=0.48\textwidth]{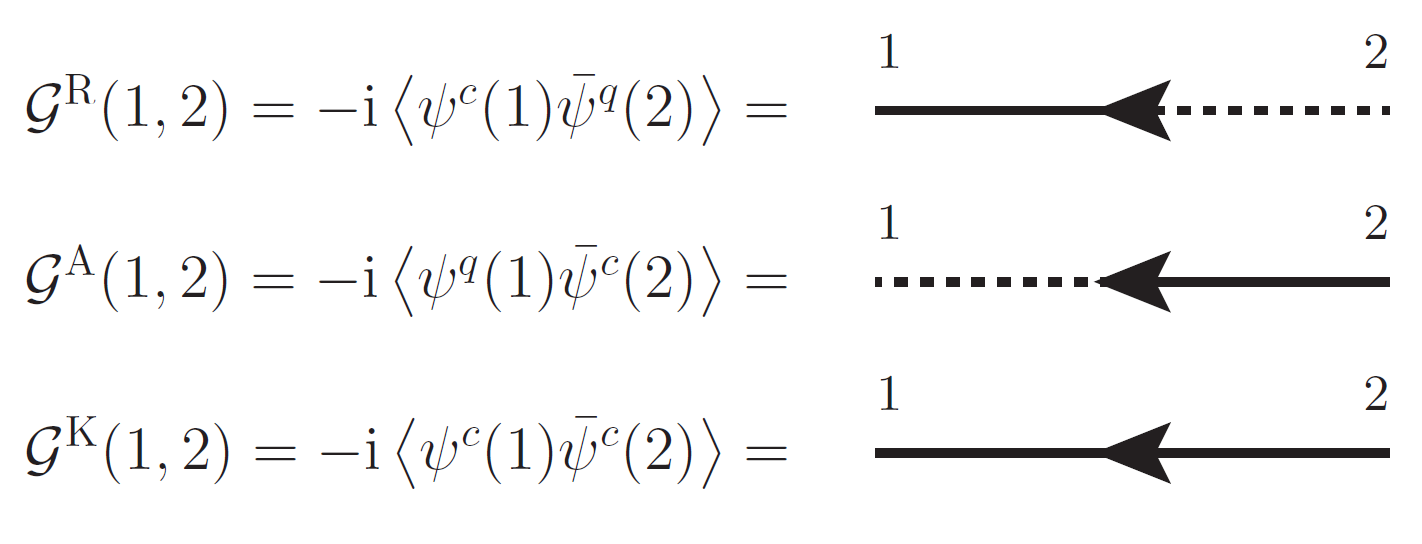}
	\caption{Noninteracting Green's functions in diagrammatic representation.}
	\label{fig:legend_green}
\end{figure}
In a diagrammatic notation, classical fields $\psi^c$ are illustrated by solid lines while quantum fields $\psi^q$ are illustrated by dashed lines. Our diagrammatic convention for the Green's function is given in Fig.~\ref{fig:legend_green}. At first order in the interaction $g(x-y)$, the corrections of the Green's functions, $\mathcal{G} = \mathcal{G}_0 + \delta\mathcal{G}$, have the following structure,
\begin{subequations}
    \label{eq:Keldysh-first-order-corrected}
	\begin{align}
	\delta\mathcal{G}^{\mathrm{R}} &\sim \mathrm{i}\int g~ \mathcal{G}_0^\mathrm{R}\mathcal{G}_0^<\mathcal{G}_0^\mathrm{R},\\
	\delta\mathcal{G}^{\mathrm{A}} &\sim \mathrm{i}\int g~ \mathcal{G}_0^\mathrm{A}\mathcal{G}_0^<\mathcal{G}_0^\mathrm{A},\\
	\delta\mathcal{G}^{\mathrm{K}} &\sim \mathrm{i}\int g~ \left[ \mathcal{G}_0^\mathrm{R}\mathcal{G}_0^<\mathcal{G}_0^\mathrm{K} + 	\mathcal{G}_0^\mathrm{K}\mathcal{G}_0^<\mathcal{G}_0^\mathrm{A} \right],
	\label{eq:Keldysh-first-order-K}
	\end{align}
\end{subequations}
where the integral is taken over the internal frequency and position variables.
The lesser Green's function $\mathcal{G}^< = \left(\mathcal{G}^\mathrm{K}+\mathcal{G}^\mathrm{A}-\mathcal{G}^\mathrm{R}\right)/2$ in place of a Keldysh Green's function is needed to cure the regularization issue for the Keldysh Green's function at equal time $\mathcal{G}^\mathrm{K}(t,t)$, as discussed in App.~\ref{sec:Regularization-perturbation}. \par

\subsection{First-order corrections to current}

Here we show how to obtain the explicit first-order correction in the interaction $g(x-y)$ to the currents $J_i = J_{0,i} + \delta J_i$ with $i=L,R$. For a better readability and in the style of Refs.~\onlinecite{Aristov_2014,Aristov_2017,Aristov_2018,Aristov_2019,nosov2020tunneling}, we use wire indices $i,j,l$ indicating the respective side of the wire. This, in principle, allows for a generalization to $N$ wires connected to reservoirs $i,j,l=1,2,...,N$ coupled to each other at the center $x<s$ where the lossy impurity is located. Moreover, we equip the interaction potential with a wire index $g_l(x-y)$ to theoretically allow for different interactions in each wire since at the end of the paper we relate our results to those obtained in a Y-junction~\cite{Aristov_2014,Aristov_2017}. \par
\begin{figure}
	\centering
	\includegraphics[width=0.49\textwidth]{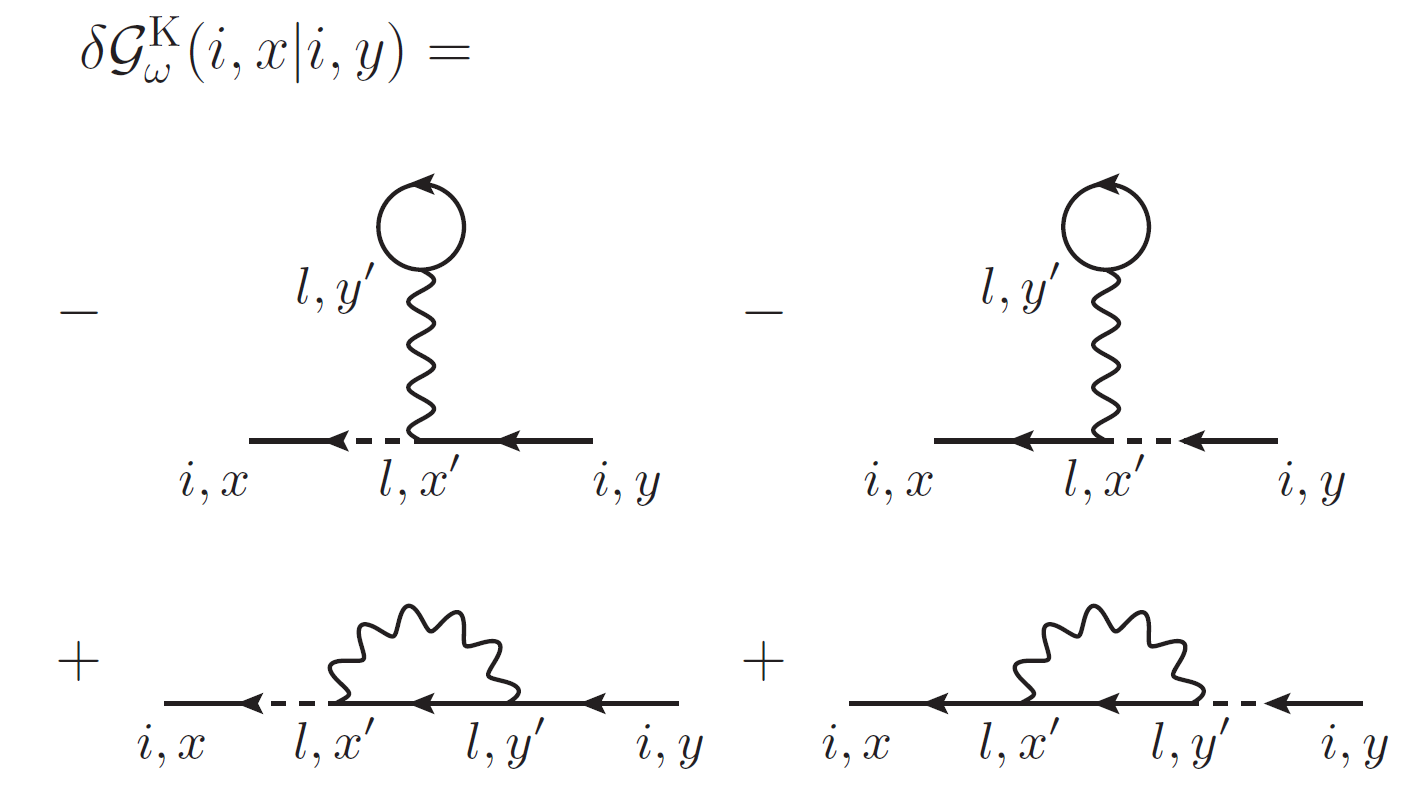}
	\caption{Keldysh diagrams corresponding to the first-order correction of the current $\delta J_i$, Eq.~\eqref{eq:Current-correction-quadratic-start}. Due to the regularization at equal times, the inner Keldysh Green's function is replaced by the lesser Green's function.}
	\label{fig:First-order-current}
\end{figure}

The average current is related to the equal-time Keldysh Green's function (or more precisely, lesser Green's function) as
\begin{equation}
J_i(x) = \frac{1}{m}\int_\omega\mathrm{Re}\left[\partial_x\mathcal{G}^<_\omega(i,x|i,y)\right]\big|_{x=y}.
\end{equation} %\marcel{why written here explicitly?}
[cf.\ Eq.~\eqref{eq:Current-definition}]. To first order in the interaction $g_l(x-y)$, the frequency integrals over the response Green's functions vanish due to causality, i.e., $\int_\omega \delta G^{\mathrm{R}}_\omega = \int_\omega \delta \mathcal{G}^\mathrm{A}_\omega =0$, such that the only contribution to the lesser Green's function $\mathcal{G}^<$ comes from the first-order correction to the Keldysh Green's function $\delta G^{\mathrm{K}}$ which is shown diagrammatically in Fig.~\ref{fig:First-order-current}. There, the interaction $g_l(x'-y')$ is represented by a wiggly line between the points $l,x'$ and $l,y'$. The first-order correction to the current $\delta J_i$ takes then the form
\begin{align}
	\nonumber & \delta J_i =
	\frac{1}{2m} \sum_l \int_{\omega,\omega',x',y'} \!\!g_l(x'-y') \\
	\nonumber & \times\text{Im}\!\left[- (\partial_x\mathcal{G}^\mathrm{R}_{0,\omega}(i,x|l,x'))\mathcal{G}^<_{0,\omega'}(l,y'|l,y')\mathcal{G}^\mathrm{K}_{0,\omega}(l,x'|i,x)\right.\\
	\nonumber & ~~~~ -\!(\partial_x\mathcal{G}^\mathrm{K}_{0,\omega}(i,x|l,x'))\mathcal{G}^<_{0,\omega'}(l,y'|l,y')\mathcal{G}^\mathrm{A}_{0,\omega}(l,x'|i,x)\\
	\nonumber & ~~~~ +\!(\partial_x\mathcal{G}^\mathrm{R}_{0,\omega}(i,x|l,x'))\mathcal{G}^<_{0,\omega'}(l,x'|l,y')\mathcal{G}^\mathrm{K}_{0,\omega}(l,y'|i,x)\\
	& ~~~~ \left. + (\partial_x\mathcal{G}^\mathrm{K}_{0,\omega}(i,x|l,x'))\mathcal{G}^<_{0,\omega'}(l,x'|l,y')\mathcal{G}^\mathrm{A}_{0,\omega}(l,y'|i,x)\right],
	\label{eq:Current-correction-quadratic-start}
\end{align}
where the first two terms correspond to the Hartree diagrams, while the second two to the Fock diagrams. These integrals can be explicitly evaluated by using the noninteracting Green's functions of Eq.~\eqref{eq:Impurity-Green} (see~App.~\ref{sec:Calculation-Details}). The computation shows that all nonvanishing contributions to the current evaluated at the end of the wires, i.e., $\delta J_i(x>L/2)$, can be collected in a regular and a singular one with respect to the voltage $V$ [cf.\ Eq.~\eqref{eq:Current-Glazman-all}], i.e., $\delta J_{i} = \delta J_{i,\text{reg.}} + \delta J_{i,\text{sing.}}$, with the latter displaying a logarithmic behavior for small voltages, namely %
\begin{align}
    \nonumber\delta J_{i,\text{sing}}&= -\sum_{j,l}\int_{k,k'>0}\!\!\left[ \tilde{g}_l(0)\!-\! \tilde{g}_l(k+k') \right]\\
    &\quad\quad\quad\quad \times\mathrm{Re}\left[\frac{\bar{S}_{ij}S_{il}\bar{S}_{ll}'f_l'S_{lj}f_j}{k-k'+\mathrm{i}0^+}\right],
    \label{eq:J^(1)_div_rewritten}
\end{align}
where $\tilde{g}(k)$ is the Fourier transform of the interaction $g(x)$. \par

\begin{figure}[t!]
	\centering
	\includegraphics[width=0.49\textwidth]{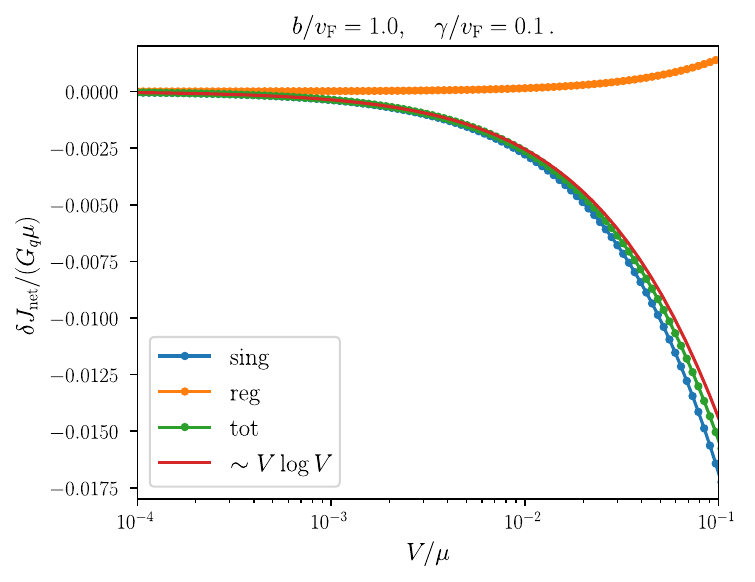}
	\caption{
    First-order corrections to the net current $\delta J_\mathrm{net} = \left(\delta J_R - \delta J_L\right)/2$. The current contributions $\delta J_\mathrm{sing}$, Eq.~\eqref{eq:J^(1)_div_rewritten}, $\delta J_\mathrm{reg}$, Eq.~\eqref{eq:Current-Glazman-all}, and $\delta J_\mathrm{tot} = \delta J_\mathrm{sing}+\delta J_\mathrm{reg}$ are integrated out numerically. A dominance of the term $\delta J_\mathrm{sing}$ over $\delta J_\mathrm{reg}$ is evident. Moreover, the results from the numerical integration give rise to the logarithmic scaling $V\ln V$ of the  analytical approximation $J_\mathrm{net} = \frac{\alpha}{2\pi}\left[\mathrm{Re}(t^2\bar{r}^2)-|rt|^2\right]V\ln|2\mu/V|$, Eq.~\eqref{eq:J-correction-net-loss}, which is plotted in comparison (red line). The negative sign indicates that the corrections for the net currents reduce the total currents for repulsive interactions $\alpha>0$, i.e., the correction flows into the opposite direction than the noninteracting net current. For the numerical integration we used $0^+\to 10^{-4}$, $g=200$, $d=0.14$ such that $\alpha \simeq 0.12$ and the parameters $b=1.0$ and $\gamma=0.1$ for the strength of the impurity.
	}
	\label{fig:Jv01g01netcontributions}
\end{figure}

To unveil the singular nature of the previous equation, we assume $T=0$,  $L \to +\infty$ in the wires, and that the momentum dependence of the interaction potential $\tilde{g}_l(k+k')$ and the scattering amplitudes $S_{jl}(k)$ are smooth near the average Fermi momentum $k_\mathrm{F}$. We then introduce the interaction parameter $\alpha_l \equiv [\tilde{g}_l(0)-\tilde{g}_l(2k_\mathrm{F})]/({2\pi v_\mathrm{F}})$. Furthermore, we denote the voltage $V_{lj} = \mu_l-\mu_j$ between reservoirs $l$ and $j$ and the average chemical potential $\mu = (\mu_L + \mu_R) / 2$. The scattering amplitudes $S_{ij}$ are evaluated at the average Fermi momentum $k_\mathrm{F}$. We find:
\begin{equation}
	\delta J_{i,\text{sing}} \simeq \frac{\alpha_l}{2\pi}\sum_{j, l} \mathrm{Re}\big[\bar{S}_{ij}S_{il}\bar{S}_{ll}S_{lj}\big]V_{lj}\ln\left\vert\frac{2\mu}{V_{lj}}\right\vert. \label{eq:Current-correction-simplification}
\end{equation}
Note that a similar formula was found in the context of Y-junctions~\cite{Aristov_2014,Aristov_2017,Aristov_2018,Aristov_2019,nosov2020tunneling}. While the singular contribution in Eq.~\eqref{eq:Current-correction-simplification} is not, strictly speaking, divergent (since it vanishes for $V \to 0$), it clearly leads to a divergent contribution in the differential conductance $G_i = \partial J_i/\partial V$, Eq.~\eqref{eq:diff_conductance}. We furthermore checked that $\delta J_{i,\text{sing}}$ represents the quantitatively largest correction to the total current: to do this, we evaluated numerically both the contributions as a function of $V$ by using a short-range interaction $g(x)=g \exp(-2|x|/d)$. The resulting voltage dependence, reported in Fig.~\ref{fig:Jv01g01netcontributions}, confirms that
$\delta J_{i,\text{sing}}$ exhibits a logarithmic scaling $V\log(V)$ and represents the leading term. %\ale{we need to adjust the plot.} 
For these reasons, in the rest of the paper we will only focus on $\delta J_{i,\text{sing}}$. \par

\subsection{Current correction in terms of scattering amplitudes}
\label{sec:Current-scattering-corrections}

Here, we provide an alternative derivation to obtain the correction to the current, based on the scattering coefficients $S_{ij}$. The latter can be derived by the first-order correction of the retarded Green's function Eq.~\eqref{eq:Impurity-Retarded-lj}~\cite{matveev1993tunneling,Yue1994,Froeml2019,Froeml2020,muller2021shape}
\begin{equation}
\delta G^{\mathrm{R}}_\omega(i,x|j,y)=\frac{1}{\mathrm{i}v_k}\delta S_{ij}(k)\ee^{\mathrm{i}k(x+y)}.
\label{eq:GR^(1)}
\end{equation}
The dominating term of the first-order correction for the scattering amplitude can be extracted from $\delta G^{\mathrm{R}}_\omega(i,x|j,y)$ and yields
\begin{multline}
	\delta S_{ij}(k) = -\frac{1}{2v_k}\sum_l\int_{k'>0} [\tilde{g}_l(0)-\tilde{g}_l(k+k')]\\	
	\times\bigg[\frac{S_{il}\bar{S}'_{ll}S_{lj}}{k-k'+\mathrm{i}0^+} + \frac{\delta_{il}S'_{ll}\delta_{lj}}{k'-k+\mathrm{i}0^+}\bigg]{f_l'},
	\label{eq:Correction-S_jl}
\end{multline}
by using the integral expression Eq.~\eqref{eq:single-k-integral}, we can extract the logarithmic scaling
\begin{multline}
    \delta S_{ij}(k) = -\sum_l \frac{\alpha_l(k)}{2} \left[ S_{il}\bar{S}_{ll}S_{lj}-\delta_{il}S_{ll}\delta_{lj}\right]\ln\left\vert\frac{k}{k-k_j}\right\vert.
    \label{eq:Correction-S_jl-compactified}
\end{multline}
In this expression, the scattering matrices are evaluated at the respective momentum $k$, i.e., $S_{ij}=S_{ij}(k)$, as well as the interaction parameter $\alpha_l(k)\equiv [\tilde{g}_l(0)-\tilde{g}_l(k+k_i)]/(2\pi v_k)$.
\par
Accordingly, the first-order correction to the scattering probability $|S_{ij}|^2$, is given by $\delta|S_{ij}|^2 = 2 \mathrm{Re}[ \bar{S}_{ij}\cdot\delta S_{ij}]$. Finally, by inserting $\delta|S_{ij}|^2$
into the generalized Landauer-Büttiker formula~\eqref{eq:Current-j}, we correctly obtain the singular correction $\delta J_{i,\text{sing}}$ to the current, Eq.~\eqref{eq:J^(1)_div_rewritten}
\begin{equation}
\delta J_{i,\text{sing}} = \int_{k>0} v_k \sum_j \delta|S_{ij}|^2 f_j.
\end{equation}
Besides providing a possibly simpler way to evaluate current contributions, this derivation also offers a more transparent interpretation of the physical processes underlying these corrections.
In fact, from Eq.~\eqref{eq:Correction-S_jl} we observe that interactions generate an additional scattering for a particle incoming from the $i$-th terminal and moving towards the $j$-th one: this scattering event is generated by the inhomogeneous density of particles in the $l$-th terminal. The factor $\propto 1/(k-k')$, at the root of the logarithmic corrections, arises from the Friedel oscillations in the particle density~\cite{matveev1993tunneling,Yue1994,Froeml2019,Froeml2020}.  
\begin{figure}[t!]
	\centering
	\includegraphics[width=0.48\textwidth]{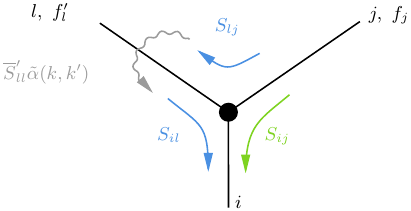}
	\caption{Visualization of the scattering events, which contribute to the current towards reservoir $i$. In the noninteracting case, the current $J_i$, Eq.~\eqref{eq:Current-j}, is carried by direct scattering from reservoir $j$ to $i$ with amplitude $S_{ij}$ (marked by green arrow). In first-order perturbation theory, the dominating current correction $\delta J_{i,\text{sing}}$, Eq.~\eqref{eq:Current-correction-simplification} is characterized by an additional rescattering event in wire $l$ with amplitude $\bar{S}_{ll}$. So fermions first scatter from $j$ to $l$  with $S_{lj}$ and then they scatter from wire $l$ to $i$ with $S_{il}$. In the sketch, we use the abbreviation $\tilde{\alpha}(k,k') = (\tilde{g}(0)-\tilde{g}(k+k')/(k-k'+\ii 0^+)$. Also note that the wire indices $i,j,l = L,R$ can coincide depending on whether we consider reflection or transmission.}
	\label{fig:Visualization-Current-correction}
\end{figure}
The physical interpretation of the current term $\delta J_{i,\text{sing}}$, Eq.~\eqref{eq:J^(1)_div_rewritten}, flowing toward reservoir $i$, is now the following [cf.\ Fig.~\ref{fig:Visualization-Current-correction}]: fermions from reservoir $j$ with momentum $k$ and distributed according to $f_j(k)$ scatter from wire $j$ to wire $l$ with amplitude $S_{lj}(k)$. Due to the interactions, in wire $l$ they scatter with the renormalized impurity, i.e., with other fermions of distribution $f_l(k')$. The corresponding scattering amplitude is proportional to $(\tilde{g}(0)-\tilde{g}(k+k'))S_{ll}(k')$ and diverges when the momenta of the interacting particles are equal $k\simeq k'$. Finally, the fermions scatters from wire $l$ to wire $i$ with amplitude $S_{il}(k)$. So instead of directly scattering from wire $j$ to wire $i$, which is the only process in the noninteracting case, there is an additional scattering event in wire $l$ between the particle and the Friedel oscillations. \par
To conclude, the singular current correction $\delta J_{i,\text{sing}}$ Eq.~\eqref{eq:J^(1)_div_rewritten} can be equivalently obtained from the first-order correction of the scattering amplitude $S_{ij}(k)$, via the generalized Landauer-B\"{u}ttiker formula \eqref{eq:Current-j}. \par

\section{Renormalization of conductances \label{sec:Renormalization-Conductances}}

In this section we show how the first order-correction of the current $\delta J_{i,\text{sing}}$, Eq.~\eqref{eq:Current-correction-simplification},  gives rise to universal logarithmic scaling which lead to a resummation of higher-order corrections by a RG treatment of differential conductances [cf.\ Eq.~\eqref{eq:diff_conductance}]. Our approach is inspired by the RG treatment of scattering amplitudes pioneered in Refs.~\onlinecite{matveev1993tunneling,Yue1994} and by the RG treatment of conductances in junctions of quantum wires~\cite{Aristov_2014,Aristov_2017,Aristov_2018,Aristov_2019,nosov2020tunneling}. \par

\subsection{Derivation of the RG equations}

As promoted before, we apply Eq.~\eqref{eq:Current-correction-simplification} to the case of a single wire (i.e., two terminals), corresponding to a $2\times 2$ scattering matrix $S(k)$ [cf.\ Eq.~\eqref{eq:Scattering-matrix}], and we assume a symmetric impurity, for which $r^L_k=r^R_k$. 
The voltage $V=\mu_R-\mu_L$ between the wires thus serves as an infrared cutoff and the average chemical potential $\mu=\left(\mu_L+\mu_R\right)/2$ is assumed to be large compared to the voltage scale. The resulting current corrections are thus
\begin{equation}
	\delta J_i \simeq \mp\frac{\alpha}{2\pi}\left[\mathrm{Re}(t^2\bar{r}^2)-|rt|^2\right]V\ln\left\vert\frac{2\mu}{V}\right\vert.
	%\delta J_\mathrm{net} &\simeq \frac{1}{2\pi}\alpha\left[\mathrm{Re}(t^2\bar{r}^2)-|rt|^2\right]V\ln\left\vert\frac{2\mu}{V}\right\vert,\\
	%\delta J_\mathrm{loss} &\simeq 0.
	\label{eq:J-correction-net-loss}
\end{equation}
We recall that here scattering amplitudes are evaluated at the average Fermi momentum, i.e., $t \equiv t_{k_\mathrm{F}}$, $r \equiv r_{k_\mathrm{F}}$. As a result, the logarithmically scaling correction to the differential conductance $\delta G$ [cf.\ Eq.~\eqref{eq:diff_conductance}] (in units of the conductance quantum $G_q={e^2}/{h}$) is given by
\begin{equation}
	\delta G  \simeq \alpha \left[\mathrm{Re}(t^2\bar{r}^2)-|rt|^2\right]\ln\left\vert\frac{2\mu}{V}\right\vert.
	%\delta G_a &\simeq g \left[\mathrm{Re}(t^2\bar{r}^2)-|rt|^2\right]\ln\left\vert\frac{2\mu}{V}\right\vert\\
	%\delta G_b &\simeq 0.
\end{equation}
From the previous equation it is evident that the perturbative approach fails for $G(V)$, since the perturbative correction $\delta G$ diverges for $V \to 0$. The logarithmic nature of the correction thus calls for a resummation of the perturbative series, which is done by means of the renormalization group~\cite{matveev1993tunneling,Yue1994,Aristov_2014,Aristov_2017,Aristov_2018,Aristov_2019,Froeml2020,muller2021shape}. \par
So far, we have only constrained the impurity to the symmetric case, i.e., $r^L=r^R$. From now on, let us additionally assume that the impurity is $\delta$-shaped, which allows us to rewrite $\mathrm{Re}(\bar{r}^2t^2)=\eta^2/2-|rt|^2$ in terms of the loss probability $\eta$ [cf.\ Eq.~\eqref{eq:eta_k-condition}]. Further using the relation $|r|^2=1-|t|^2-\eta$, the first-order conductance $\delta G$ can thus be uniquely related to the zeroth-order expression $G_0= |t|^2+\eta/2$ as
\begin{align}
		\delta G &= 2\alpha \left[ \frac{\eta}{2} - G_0\left(1-G_0\right)\right]\ln\left\vert\frac{2\mu}{V}\right\vert.
		\label{eq:Ga-Gb-dissipative}
\end{align} \par
Let us comment on how the loss probability $\eta$ affects the transport properties. As the two corrections $\delta J_L$ and $\delta J_R$, Eq.~\eqref{eq:J-correction-net-loss}, come with an opposite sign, the correction to the loss current $\delta J_\mathrm{loss} = -\delta J_L-\delta J_R$ [cf.\ Eq.~\eqref{eq:non-interacting:J_loss}], does not exhibit a logarithmic scaling with respect to the voltage $V$.  Thus in context of transport properties, the loss probability $\eta$ appearing in Eq.~\eqref{eq:Ga-Gb-dissipative} for the conductance can be considered as an input parameter. In the limit $V\to 0$, we finally derive the RG equation of the scale-dependent conductance $G(\Lambda)$ with logarithmic RG flow parameter $\Lambda = \ln\vert{2\mu}/{V}\vert$
\begin{equation}
	    \frac{\mathrm{d} G}{\mathrm{d} \Lambda} = 2\alpha \left[\frac{\eta}{2}-G(1-G)\right].
		\label{eq:RG-equations}
\end{equation}
This equation is one of the central results of our work and it generalizes the results of Refs.~\onlinecite{matveev1993tunneling,Yue1994} to the case of wires with localized particle loss. \par
In previous works~\cite{Froeml2019,Froeml2020,muller2021shape}, the RG analysis was set up in terms of the perturbative corrections of the scattering parameters, whose corrections scale logarithmically near the Fermi surface, i.e., $\delta S_k\sim\ln|k-k_\mathrm{F}|$ [cf.\ Eq.~\eqref{eq:Correction-S_jl-compactified}]. On the contrary, transport phenomena are described by a logarithmic scaling with respect to the voltage $\sim \ln|V|$ [cf.\ Eqs.~\eqref{eq:J-correction-net-loss} and \eqref{eq:Ga-Gb-dissipative}]. Inserting the corrections of the scattering parameters (scaling with $\ln|k-k_\mathrm{F}|$) into Eq.~\eqref{eq:G_L,R(t_L,R} does not yield the correction of the conductance $\delta G$, Eq.~\eqref{eq:Ga-Gb-dissipative}, (scaling with $\ln |V|$) [see also discussion in App.~\ref{sec:RG-scattering}]. It is crucial that $V\to 0$ is only taken when initializing the RG equation \eqref{eq:RG-equations}. This is because the loss current, which is mainly carried by contributions from the bulk [cf.\ Fig.~\ref{fig:transport-processes}], does not exhibit a singular correction and thus provides $\eta$ as an additional parameter in the RG flow of the conductance. \par

\subsection{Solutions of the RG equation}

\begin{figure}[t!]
	\centering
	\includegraphics[width=0.45\textwidth]{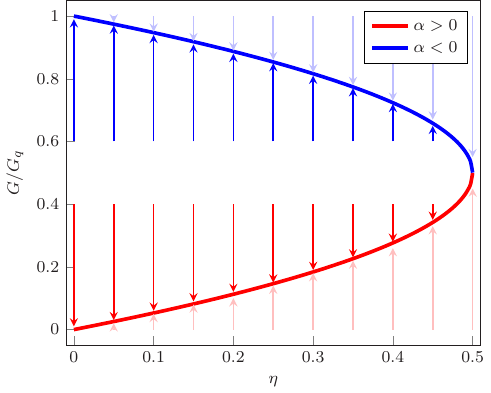}
	\caption{Visualization of the RG flow for the conductance $G$ in a two-lead wire including localized particle loss: For a $\delta$ impurity, the allowed range for conductances is within the parabola $G(1-G)=\eta/2$ which is bounded by the fixed points $G^*$ and corresponds to purely dissipative impurities. For repulsive interactions $\alpha>0$ the conductance is decreased to the lower red branch of the parabola whereas for attractive interactions $\alpha<0$ the conductance is enhanced to the upper blue branch of the parabola. The Kane-Fisher result is only retrieved in the dissipationless case, i.e., $\eta=0$.}
	\label{fig:rgflowgagbtwoleadloss}
\end{figure}

\begin{figure*}[t!]
	{\centering
	{\includegraphics[width=0.45\textwidth]{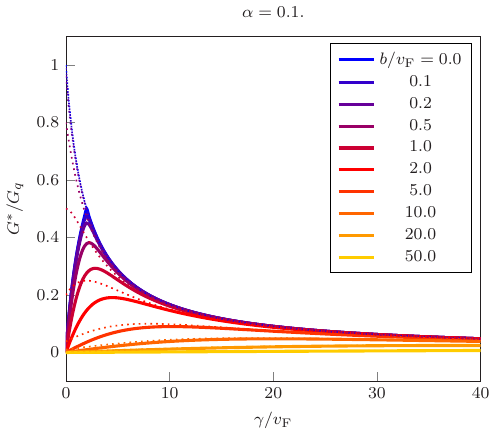}
    \includegraphics[width=0.45\textwidth]{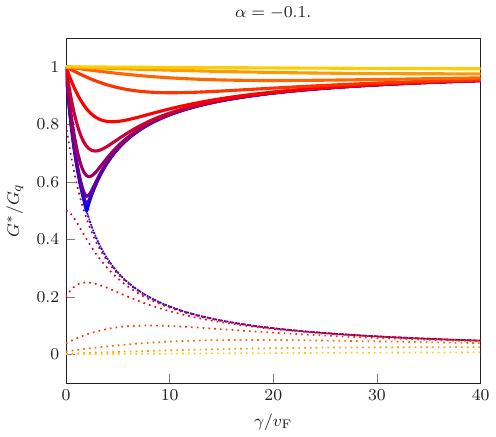}}
	\caption{Fixed points $G_\pm^*$ (solid lines) for the conductances in units of the conductance quantum $G_q$ dependent on the dissipation strength $\gamma$ for different coherent strengths $b$ for repulsive (left panel) and attractive (right panel) interactions. The corresponding noninteracting values $G_0$ of the conductances are shown as thin dotted lines. The nature of the of fluctuation-induced quantum Zeno effect is represented in the sense that for large dissipation strength $\gamma$, the conductance is lowered (repulsive case) or enhanced (attractive case), however ideal reflectivity and transparency are only retrieved for $\gamma=0$.}\label{fig:dissconductancefp}}
\end{figure*}

The RG equation~\eqref{eq:RG-equations} is exactly solvable [cf.\ Eq.~\eqref{eq:G(Lambda)_exact}], and its flow is depicted in Fig.~\ref{fig:rgflowgagbtwoleadloss}.
It predicts two lines of fixed points depending on the bare loss probability $\eta$
\begin{equation}
	G^*_\pm = \frac{1}{2}\left(1\mp\sqrt{1-2\eta}\right).
\end{equation}
This is in stark contrast with the result for coherent impurities, where only two fixed points are allowed, namely $G^*= 0,1$. Due to the dissipation, the loss probability $\eta$ is an additional marginal parameter in the RG analysis. The fixed points are thus a function of $\eta$, which is shown in Fig.~\ref{fig:rgflowgagbtwoleadloss}. The initial conditions of the RG flow are given by the noninteracting conductance $G(\Lambda=0)=G_0=|t|^2+\eta/2$. The conditions for a scattering matrix of a $\delta$-shaped impurity [cf.\ Eq.~\eqref{eq:Scattering-delta}] lead to the boundary condition $G(1-G)\geq\eta/2$ [cf.\ Eq.~\eqref{eq:Inequality_G}] which is exactly bound by the manifold of the fixed points $G^*$, and is valid for the whole RG flow $G(\Lambda)$. Thus, the RG flow reduces the conductance $G$ for repulsive interactions $\alpha>0$ to the minimally allowed value and enhances it for attractive interactions $\alpha<0$ to the maximally allowed value for a certain loss probability $\eta$ [cf.\ Fig.~\ref{fig:dissconductancefp}]. However, ideal reflectivity $G^* = 0$ and transparency $G^* =1$, which are found in the Kane-Fisher problem~\cite{kane1992transmission,kane1992transport}, are only reached in the case of a coherent impurity where $\eta = 0$ [cf.\ Fig.~\ref{fig:rgflowgagbtwoleadloss}]. \par
\par
From the conditions of the $\delta$ impurity (cf.\ Eq.~\eqref{eq:Scattering-delta}), we know that the equality $G_0(1-G_0)=\eta/2$ is valid in the case of a purely dissipative impurity, where the scattering amplitudes are real-valued [cf.\ Eq.~\eqref{eq:Scattering-delta}]. In this case, the conductance $G_0$ already coincides with the fixed point $G^*$ and interactions $g$ do not renormalize it anymore. Interpreting our results as a renormalization of the impurity, one can state that interactions maximize the dissipation in such a way that the system is only affected by the dissipation $\gamma$ and not by the coherent barrier strength $b$. Indeed, the conductance fixed points $G^*$, which are driven by the particle interactions $g$ exhibit a nonmonotonic behavior with respect to the dissipation strength $\gamma$ [cf.\ Fig.~\ref{fig:dissconductancefp}]. This is a new incarnation of the fluctuation-induced quantum Zeno effect~\cite{Froeml2019,Froeml2020}. We suggest that such a dependence can be measured in experimental setups like ultracold fermionic atoms in an optically shaped quantum point contact~\cite{lebrat2019local, corman2019quantized}. \par
We derive a new nonuniversal conductance scaling by linearizing the RG equations \eqref{eq:RG-equations} around the fixed point, i.e., $G = G^*_\pm + \delta G$
\begin{subequations}
	\begin{align}
	\frac{\mathrm{d} \delta G}{\mathrm{d}\Lambda} &= -2|\alpha| \sqrt{1-2\eta}~\delta G\\
	\Rightarrow \delta G &\propto |V|^{2|\alpha|\sqrt{1-2\eta}}.
	\end{align}
	\label{eq:Power-Law}
\end{subequations}
Note that the parameter space of delta impurities restricts the loss probability to $\eta<1/2$ such that the nonuniversal scaling exponent $2|\alpha|\sqrt{1-2\eta}$ stays well defined~\cite{muller2021shape}.\par
In comparison to the conductance scaling in the Kane-Fisher problem~\cite{kane1992transmission,kane1992transport}, the conductance $G$ of the dissipative wire now also depends on the loss probability $\eta$, which is an additional marginal parameter in the RG flow, and therefore on the microscopic values of the coherent strength $b$ and dissipation strength $\gamma$ of the impurity. \par
This result provides an additional perspective to previously considered setups~\cite{Froeml2019,Froeml2020, muller2021shape} without explicit external particle reservoirs. There, we neglected the nonequilibrium effect of a finite voltage $V$ and calculated the renormalization of the lossy impurity directly on the level of the scattering probabilities $|t|^2$ and $|r|^2$ at the Fermi momentum. However, here the focus is on measurable transport properties, which in the dissipative case, also depend on states inside the Fermi sea. That is why the dissipative conductance scales differently than predicted by the scattering amplitudes evaluated at the Fermi momentum. The multiplicative form of the exponent is unusual and comes from the fact that the dissipation strength acts both as a marginal coupling in the RG flow and as a cutoff scale. In that sense, we find agreement with Ref.~\onlinecite{muller2021shape}. We may gain understanding of that finding by describing the particle loss by a Y-junction with one empty reservoir as discussed in the next section.  

\subsection{Comparison with Y-junctions}

\begin{figure}[t!]
	\centering
	\includegraphics[width=0.48\textwidth]{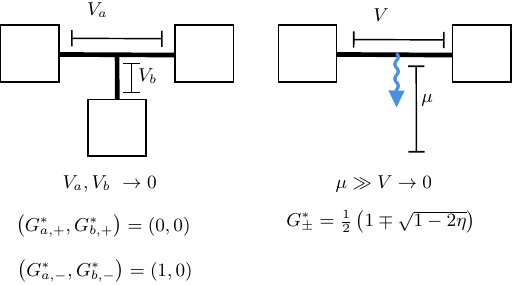}
	\caption{Relation between the Y-junction and the dissipative quantum wire. In the limit of two small voltages $V_a, V_b \to 0$ of a Y-junction, we end up in the one-dimensional system where the Kane-Fisher fixed points of ideal reflectivity and transparency are found. In the limit of $V_b\gg V_a$, we arrive at the dissipative wire between two leads. There we find the RG flow of a dissipative conductance where ideal reflectivity and transparency are not reached anymore.}
	\label{fig:Conclusion-comparison}
\end{figure}

In this section, we briefly comment how our results are related to those obtained in a junction of three wires coupled to reservoirs with chemical potentials $\mu_1$, $\mu_2$ and $\mu_3$, a so-called Y-junction (cf.\ Fig.~\ref{fig:Conclusion-comparison}). Using the bosonization technique, junctions of Luttinger liquids have been studied at arbitrary interaction strength in the limit of weak hopping between the wires~\cite{Nayak1999Resonant}. As we are mainly interested in the effect of particle loss, we follow Refs.~\onlinecite{Aristov_2014,Aristov_2017,Aristov_2018,Aristov_2019}, where junctions out of equilibrium are studied in the limit of weak interactions. \par
In the noninteracting case, the transport through a dissipative wire can be identified with the transport through a Y-junction \cite{Uchino2022Comparative}. This is because the noninteracting net current $J_\mathrm{net}=(J_R-J_L)/2$ is only carried by states at the Fermi level [cf.\ Sec.~\ref{sec:Non-interacting}]. When interactions are included, however, scattering events with the dissipation-induced holes inside the bulk (cf.\ Fig.~\ref{fig:transport-processes}) renormalize the conductance in a way which depends on the chemical potential $\mu$, i.e., the voltage drop between the wire and the environment. We will argue that the identification between a dissipative wire and a Y-junction is valid in the limit $\mu_3\ll\mu_1,\mu_2$. \par
We refer to the wire between reservoirs with $\mu_1$ and $\mu_2$ as the main wire which can be identified with our dissipative wire between the reservoirs $\mu_L$ and $\mu_R$. The third wire toward the reservoir with $\mu_3$, also considered as secondary wire, has been absent in our consideration. Instead of this, we used the theory of an open quantum system. However, our loss probability $\eta$ can be identified with the transmission probability $|\tau|^2$ between the main wire and the secondary wire. Whereas the Y-junction is characterized by a unitary $3\times 3$-scattering matrix $S(k)$, in our case of a dissipative wire we consider nonunitary dynamics where particles may only leave the system at the impurity, but not enter such that we have a nonunitary $2\times 2$-scattering matrix [cf.\ Eq.~\eqref{eq:Scattering-matrix}].
In a Y-junction it is useful to express the transport properties with respect to the voltage $V_a = \mu_1-\mu_2$ across the main wire and the voltage $V_b = \left(\mu_1+\mu_2\right)/2-\mu_3$ across the secondary wire. With this consideration it makes sense to define two different conductances, namely $G_a = \partial(J_2-J_1)/\partial (2V_a)$ along the main wire and $G_b = \partial J_3/\partial V_b$ along the secondary wire. The corresponding noninteracting values for the currents $J_i$, Eq.~\eqref{eq:Current-j}, as well as the logarithmically scaling corrections $\delta J_i$, Eq.~\eqref{eq:Current-correction-simplification}, can be determined within the formalism developed above as it can be extended to a wire index $i=1,2,3$ (cf.\ App.~\ref{sec:Y-junction}).\par
The interplay of different voltage scales $V_a$ and $V_b$ makes it ambiguous how to arrange a cutoff scale $\Lambda$ to set up a useful RG equation. Taking the limit of $V_a, V_b \to 0$ where a single cutoff scale $\Lambda$ can be identified again~\cite{Aristov_2017,Aristov_2018,Aristov_2019}, and assuming vanishing interactions in the secondary wire, $\alpha_3=0$, reproduces the universal fixed points $(G^*_a, G^*_b) = (0,0)$ and $(1,0)$ of ideal reflectivity and ideal transparency from the Kane-Fisher problem~\cite{kane1992transport,kane1992transmission}. Hence, the secondary wire does not qualitatively alter the conduction properties of the primary wire through a potential barrier.\par 
However, here we consider a different limit: a dissipative wire is realized in the limit of $V = V_a\ll V_b =\mu$, i.e., $\mu_3\to 0$. Here the average chemical potential $\mu$ of the main wire introduces an additional macroscopic scale which is fixed and thus only one scale parameter $\Lambda= \ln|2\mu/V|$ remains logarithmically divergent for small voltages $V$. In this physical scenario, there is only one scaling conductance $G_a$ left which can be identified with the dissipative conductance $G$ discussed before. Besides that, the macroscopic chemical potential difference ensures the third wire to be effectively noninteracting due to the absence of particles, leaving the dissipation Markovian. We know from the previous section that in this scenario unitarity of the reduced scattering matrix is lost because the loss probability, as an additional nonuniversal parameter, prevents the RG flow from reaching the case of ideal reflectivity and transparency, respectively. This is the reason why in our consideration of a dissipative quantum wire we end up in a new nonuniversal power law for the dissipative conductance $G$ [cf.\ Eq.~\eqref{eq:Power-Law}]. The problem still exhibitis scaling, but the scaling exponent depends on a microscopic parameter of the system, $\eta$. This is similar to the behavior found in critical phases, such as the Berezinskii-Kosterlitz-Thouless phase in low temperature systems with $U(1)$ symmetry~\cite{Kosterlitz:1973xp}. \par
We conclude that a Y-junction where one of the reservoirs is kept empty acts just like a two-terminal junction with a localized particle loss in terms of the conduction properties. Hence, such a Y-junction is a potential way of experimentally realizing a lossy impurity and measuring conductance through such a conductance. This might be the way to go in solid state setups, while in ultracold atoms the setup described with two reservoirs and a loss induced by other mechanisms may be more natural. An overview of the relation between a Y-junction and the dissipative wire with a finite voltage is given in Fig.~\ref{fig:Conclusion-comparison}.

\section{Conclusion}

In this article, we develop a theory to compute observables in a quantum wire featuring a localized particle loss and attached to imbalanced reservoirs. We distinguish between the loss current, which gives rise to the particles moving out of the system, and the net current through the wire. \par
Whereas in previous works~\cite{Froeml2019,Froeml2020} we analyzed scattering properties, here, our focus is on the transport properties through a lossy impurity including imbalanced reservoirs using a RG scheme in terms of the voltage. In presence of particle loss, the RG scaling behavior of the conductance cannot be solely understood in terms of the system's scattering properties at the Fermi edge since the loss current is also carried by states far below the Fermi energy. When interactions are taken into account, this affects the net current as well. We find a logarithmically scaling behavior for the net current at first-order to the interactions in terms of the voltage, but not for the loss current. As a result, the conductance of the dissipative wire scales logarithmically with respect to the voltage, but the loss probability appears as an additional marginal parameter to the conductance's RG flow. Consequently, the conductance's fixed points depend on the loss probability and the renormalized values inherit information on the strength and shape of the potential in addition to the strength of the interactions. \par
In particular, the conductance never renormalizes to zero (ideal reflectivity for repulsive interactions) or to the conductance quantum (ideal transparancy for attractive interactions), but rather to finite values, which depend nonmonotonically on the strength of the impurity, realizing an incarnation of the fluctuation-induced quantum Zeno effect. Moreover, we find that the conductance's dependence on the interaction occurs with a different power law compared to the equilibrium case, which not only depends on the strength of the interactions, but also on the bare loss probability. This is similar to the modified power law of a dissipative impurity of a finite size~\cite{muller2021shape}. We conclude that the combination of localized particle loss and finite parameters like voltage or impurity shape modifies the universal scaling properties around ideal reflectivity and transparency, which was retrieved in previous works~\cite{kane1992transport,kane1992transmission,matveev1993tunneling,Yue1994,Froeml2019,Froeml2020}. \par
Additionally, the localized particle loss in a quantum wire can also be mimicked by a third reservoir with a large voltage drop. Our results can be retrieved by the conductance scaling in a so-called Y-junction, i.e., the connection of wires coupled to three reservoirs~\cite{Aristov_2014,Aristov_2017,Aristov_2018,Aristov_2019,nosov2020tunneling}, when one of the voltages is kept large leading to a more nonuniversal behavior.\par
On the experimental side, our results provide concrete predictions, such as the impact of the impurity strength on the exponent in the scaling of the conductance through the impurity or the value of the conductance for different interaction strengths. We suggest that our theory could be tested in current experimental setups with cold atoms~\cite{Brantut2012,Krinner2015,krinner2017two,lebrat2019local,Husmann2018,Krinner8144,Visuri2023,Huang2023,fabritius2024irreversible}. Furthermore, our results confirm that the presence of the localized particle loss modifies the critical properties of the corresponding equilibrium theory, implying that the setting of lossy impurities in quantum wires is an ideal setting to study quantum critical behavior out of equilibrium that is clearly different from its equilibrium counterpart. 

~\par

\begin{acknowledgments}
	We acknowledge support by the funding from the Deutsche Forschungsgemeinschaft (German Research Foundation) under Germany’s Excellence Strategy–Cluster of Excellence Matter and Light for Quantum Computing (ML4Q) Grant No. EXC 2004/1 390534769, by the Deutsche Forschungsgemeinschaft Collaborative Research Center (CRC) 1238 Project No. 277146847–project C04. M.G. acknowledges funding from the International Max Planck Research School for Quantum Science and Technology (IMPRS-QST).
\end{acknowledgments}

\appendix

\section{Simplification of regularization issues inside the Green's functions \label{sec:Regularization-Green}}

In the absence of interactions and the impurity ($g, b, \gamma = 0$), the Hamiltonian, Eq.~\eqref{eq:Hamiltonian}, takes the simple form $\hat{H}_0=\int\mathrm{d}x~ \hat{\psi}^\dagger\frac{\partial_x}{2m}\hat{\psi}$ and the fermionic operators are stationary, i.e., $\hat{\psi}(t) = \ee^{-\mathrm{i}k^2/(2m)(t-t')}\hat{\psi}(t')$. The corresponding noninteracting Green's functions are then obtained as:
\begin{widetext}
	\begin{subequations}
		\begin{align}
		\mathcal{G}^\mathrm{R}_{0,\omega}(x,y) &= \int_t \ee^{\mathrm{i}\omega t} (-\mathrm{i})\theta(t)\left\langle\left\{\hat{\psi}(t,x),\hat{\psi}^\dagger(0,y)\right\}\right\rangle =  \int_k \frac{\ee^{\mathrm{i}k(x-y)}}{\omega+\mathrm{i}0^+-k^2/(2m)}\simeq \frac{\sqrt{2m}}{2\mathrm{i}}\frac{\ee^{\mathrm{i}\sqrt{2m\omega}|x-y|}}{\sqrt{\omega}},\\
		\nonumber \mathcal{G}^\mathrm{K}_{0,\omega}(x,y) &= \int_t \ee^{\mathrm{i}\omega t} (-\mathrm{i})\left\langle\left[\hat{\psi}(t,x),\hat{\psi}^\dagger(0,y)\right]\right\rangle\\
		\nonumber &= \int_k \ee^{\mathrm{i}k(x-y)}\frac{-2\mathrm{i}0^+}{(\omega\!-\! k^2/(2m))^2+(0^+)^2}\left[\theta^+(k)h_L(k^2/(2m))\!+\!\theta^+(-k)h_R(k^2/(2m))\right]\\
		&\simeq \int_k \ee^{\mathrm{i}k(x-y)}(-2\pi\mathrm{i})\delta(\omega\!-\! k^2/(2m))\left[\theta^+(k)h_L(k^2/(2m))\!+\!\theta^+(-k)h_R(k^2/(2m))\right] .		
		\end{align}
	\end{subequations}
	\label{eq:Green_start}
\end{widetext}
In the last step, we neglect the regularization $0^+$ which results from solving the time integrals. The fermionic distribution functions are defined as $h_i(\omega)=1-2f_i(\omega)$ and originate from the expectation values in momentum space $\left\langle\hat{\psi}^\dagger(k)\hat{\psi}(k')\right\rangle = \delta_{kk'}\left[\theta^+(k)f_L(k^2/(2m))+\theta^+(-k)f_R(k^2/(2m))\right]$ with the strict Heaviside function $\theta^+(k>0)=1$, $\theta^+(k\leq 0)=0$. This means, fermions with a positive (respectively, negative) momentum come from the left (respectively, right) reservoir with chemical potential $\mu_L$ (respectively, $\mu_R$).\par
In the stationary state, the particle density $n$ and current $J$ are formulated as frequency integrals of the lesser Green's function $\mathcal{G}^<_\omega(x,y) = -\mathrm{i}\left\langle\hat{\psi}^\dagger(\omega,y)\hat{\psi}(\omega,x)\right\rangle$:
\begin{align}
	\nonumber n(x) &= \left\langle\hat{\psi}^\dagger(t,x)\hat{\psi}(t,x)\right\rangle = \int_\omega \left\langle\hat{\psi}^\dagger(\omega,x)\hat{\psi}(\omega,x)\right\rangle\\
	&= -\mathrm{i}\int_\omega \mathcal{G}^<_\omega(x,x),
	\label{eq:Density-definition}\\
	\nonumber J(x) &= \frac{1}{m}\mathrm{Im}\left\langle\hat{\psi}^\dagger(t,x)\partial_x\hat{\psi}(t,x)\right\rangle\\
	\nonumber &= \frac{1}{m}\mathrm{Im}\int_\omega\left\langle\hat{\psi}^\dagger(\omega,x)\partial_x\hat{\psi}(\omega,x)\right\rangle\\
	&= \frac{1}{m}\int_\omega\mathrm{Re}\left.\left(\partial_x\mathcal{G}^<_\omega(x,y)\right)\right\vert_{x=y}.
	\label{eq:Current-definition}
\end{align}
The lesser Green's function fulfills the identity $\mathcal{G}^< = \frac{1}{2}\left(\mathcal{G}^\mathrm{K}+\mathcal{G}^\mathrm{A}-\mathcal{G}^\mathrm{R}\right)$. As the difference of the response Green's functions,
\begin{align}
    (\mathcal{G}_0^\mathrm{A}-\mathcal{G}_0^\mathrm{R})_\omega(x,y)\simeq \int_k \ee^{\mathrm{i}k(x-y)}2\pi\mathrm{i}~\delta(\omega-k^2/(2m)),
\end{align}
and the Keldysh Green's function $\mathcal{G}^\mathrm{K}$ contain a momentum integral over $\delta(\omega-k^2/(2m))$ the relevant frequencies are automatically related to the on-shell condition $\omega = k^2/(2m)$ and therefore are reduced to nonnegative values. Consequently, the frequency integral over the Keldysh Green's function can be rewritten as:
\begin{align}
\nonumber &\int_\omega \mathcal{G}^\mathrm{K}_{0,\omega}(x,y)\\
\nonumber &\simeq -\mathrm{i}\int_{k>0}\!\!\left[\ee^{\mathrm{i}k(x-y)}h_L(k^2/(2m))\!+\!\ee^{-\mathrm{i}k(x-y)}h_R(k^2/(2m))\right]\\
&=\!\int_{\omega>0}\!\!\frac{\sqrt{2m}}{2\mathrm{i}\sqrt{\omega}}\!\left[\ee^{\mathrm{i}\sqrt{2m\omega}(x-y)}h_L(\omega)\!+\!\ee^{-\mathrm{i}\sqrt{2m\omega}(x-y)}h_R(\omega)\right].
\label{eq:integral-K}
\end{align}
By use of the momentum $k=\sqrt{2m\omega}$ and velocity $v_k = k/m$, let us rewrite the Green's functions as:
\begin{subequations}
    \label{eq:Green-unperturbed-before}
	\begin{align}
		\mathcal{G}^\mathrm{R}_{0,\omega}(x,y) &= \frac{1}{\mathrm{i}v_k}\ee^{\mathrm{i}k|x-y|},\\
		%(\mathcal{G}_0^\mathrm{A})_\omega(x,y) &= \overline{(\mathcal{G}^\mathrm{R}_0)_\omega(y,x)},\\
		\mathcal{G}^\mathrm{K}_{0,\omega}(x,y) &= \frac{1}{\mathrm{i}v_k}\left[\ee^{\mathrm{i}k(x-y)}h_L\!+\!\ee^{-\mathrm{i}k(x-y)}h_R\right].
	\end{align}
\end{subequations}
In these expressions, the frequencies $\omega$ are restricted to positive values. \par
Finally, it is useful to change into the notation where the position variables on the two sides of the impurity are restricted to positive values $x>0$ [cf.\ Eq.~\eqref{eq:LR-basis}] and the Green's functions $\mathcal{G}^\mathrm{R,A,K}_{0,\omega}(x,y)$ of Eq.~\eqref{eq:Green_start} are translated to $2\times 2$-matrix expressions $\mathcal{G}^{\mathrm{R,A,K}}_{0,\omega}(i,x|j,y)$ depending on which side of the impurity $i,j=L,R$ the two position variables $x,y$ belong to:
\begin{subequations}
	\begin{align}
	\mathcal{G}^\mathrm{R}_{0,\omega}(i,x|j,y) &= \frac{1}{\mathrm{i}v_k}\left[\delta_{ij}\ee^{\mathrm{i}k|x-y|}+\sigma_{ij}\ee^{\mathrm{i}k(x+y)}\right],\\
	%(\mathcal{G}_0^\mathrm{A})_\omega(x,y) &= \overline{(\mathcal{G}^\mathrm{R}_0)_\omega(y,x)},\\
	\nonumber\mathcal{G}^\mathrm{K}_{0,\omega}(i,x|j,y) &= \frac{1}{\mathrm{i}v_k}\left[\delta_{ij}h_j\ee^{-\mathrm{i}k(x-y)}\!+\!\sigma_{ji}^1h_i\ee^{-\mathrm{i}k(x+y)}\right.\\
	&\left.+ \sigma_{ij}^1h_j\ee^{\mathrm{i}k(x+y)}+\sum_l\sigma_{jl}^1\sigma_{il}^1h_l\ee^{\mathrm{i}k(x-y)}\right],
	%		(\mathcal{G}_0^<)_\omega(x,y) &= -\frac{1}{\mathrm{i}v_k}\left[\ee^{\mathrm{i}\sqrt{2m\omega}(x-y)}f_1\!+\!\ee^{-\mathrm{i}\sqrt{2m\omega}(x-y)}f_2\right].
	\end{align}
	\label{eq:Green-unperturbed}
\end{subequations}
with the $x$-Pauli matrix $\sigma^1$.

\section{Dyson equations for impurity action \label{sec:Dyson}}

The contribution of the impurity to the Keldysh action  $S_\mathrm{imp}$ is calculated by a Trotter decomposition of the quantum master equation \eqref{eq:Lindblad-equation}~\cite{sieberer2016keldysh,sieberer2023universality} and takes the following form:
\begin{align}
S_\mathrm{imp} &= \int_{x,t}\left(\bar{\psi}^c,\bar{\psi}^q\right)\left(\begin{matrix}
0 & -\bar{u}(x) \\ -u(x) & \mathrm{i}\gamma(x) \end{matrix}\right) \left(\begin{matrix}
\psi^c \\ \psi^q
\end{matrix}\right),
\label{eq:Impurity-action}
\end{align}
with the impurity potential function defined as $u(x) = b(x) - \frac{\mathrm{i}}{2} \gamma(x)$. Whenever the Keldysh action takes the quadratic form:
\begin{align}
	S = \int_{\omega,x,y} \left(\bar{\psi}^c,\bar{\psi}^q\right)_{\omega,x}\left(\begin{matrix}
	0 & P^\mathrm{A}\\ P^\mathrm{R} & P^\mathrm{K} \end{matrix}\right)_{\omega,x,y}\left(\begin{matrix}
	\psi^c \\ \psi^q
	\end{matrix}\right)_{\omega,y},
\end{align}
the corresponding Green's functions are often exactly solvable. The matrix $P$ inside the action is the inverse of the matrix of Green's functions $\mathcal{G}$~\cite{kamenev_2011}, i.e.:
\begin{align}
	P = \left(\begin{matrix}
	0 & P^\mathrm{A}\\ P^\mathrm{R} & P^\mathrm{K} \end{matrix}\right) = \left(\begin{matrix}
	\mathcal{G}^\mathrm{K} & \mathcal{G}^\mathrm{R}\\ \mathcal{G}^\mathrm{A} & 0 \end{matrix}\right)^{-1} = \mathcal{G}^{-1}.
\end{align}
Note that the advanced Green's function is the adjoint of the retarded one, i.e., $\mathcal{G}^\mathrm{A}_{\omega}(x,y) = [\mathcal{G}^\mathrm{R}_\omega(y,x)]^*$. We start from the Green's functions $\mathcal{G}_0^\mathrm{R},\mathcal{G}_0^\mathrm{A},\mathcal{G}_0^\mathrm{K}$ in Eqs.~\eqref{eq:Green-unperturbed} belonging to the unperturbed quadratic Hamiltonian $\hat{H}_0$ and thus to the action components $P_0^\mathrm{R},P_0^\mathrm{A},P_0^\mathrm{K}$, i.e., $P_0 \cdot \mathcal{G}_0 = \mathds{1}$. The impact of the additional action term $S_\mathrm{imp}$, evoked by the impurity with the components $P_\mathrm{imp}^\mathrm{R}=-u=(P_\mathrm{imp}^\mathrm{A})^\dagger$ and $P_\mathrm{imp}^\mathrm{K}=\mathrm{i}\gamma$ [cf.\ Eq.~\eqref{eq:Impurity-action}], onto the Green's functions can be taken into account by the following Dyson equations:
\begin{subequations}
	\begin{align}
		\mathds{1} &= (P_{0}^\mathrm{R}+P_\mathrm{imp}^\mathrm{R})\mathcal{G}^\mathrm{R} \Rightarrow \mathcal{G}^\mathrm{R} = \mathcal{G}_0^\mathrm{R} - \mathcal{G}_0^\mathrm{R}P_\mathrm{imp}^\mathrm{R}\mathcal{G}^\mathrm{R},\\
		\mathds{1} &= (P_{0}^\mathrm{A}+P_\mathrm{imp}^\mathrm{A})\mathcal{G}^\mathrm{A} \Rightarrow \mathcal{G}^\mathrm{A} = \mathcal{G}_0^\mathrm{A}-\mathcal{G}_0^\mathrm{A}P_\mathrm{imp}^\mathrm{A}\mathcal{G}^\mathrm{A},\\
		\nonumber \mathcal{G}^\mathrm{K} &= - \mathcal{G}^\mathrm{R}P^\mathrm{K}\mathcal{G}^\mathrm{A}\\
		\nonumber &= \mathcal{G}_0^\mathrm{K}-\mathcal{G}^\mathrm{R}P_\mathrm{imp}^\mathrm{K}\mathcal{G}^\mathrm{A}-\mathcal{G}^\mathrm{R}P_\mathrm{imp}^\mathrm{R}\mathcal{G}_0^\mathrm{K}-\mathcal{G}_0^\mathrm{K}P^\mathrm{A}_\mathrm{imp}\mathcal{G}^\mathrm{A}\\
		&~~~~ +\mathcal{G}^\mathrm{R}P_\mathrm{imp}^\mathrm{R}\mathcal{G}_0^\mathrm{K}P_\mathrm{imp}^\mathrm{A}\mathcal{G}^\mathrm{A}.
	\end{align}
	\label{eq:Dyson}
\end{subequations}\par
As an example for an asymmetric impurity, we explicitly calculated the impurity Green's functions for to different delta peaks with $u(x) = u_L\delta(x+s)+u_R\delta(x-s)$ and $\gamma(x) = \gamma_L\delta(x+s)+\gamma_R\delta(x-s)$. In the notation with two wires $i = L,R$, this is translated to potential terms $u_i(x)=u_i\delta(x-s)$ and $\gamma_i(x)=\gamma_i\delta(x-s)$. In this particular case, the Dyson equations \eqref{eq:Dyson} are exactly solvable by the following matrix expressions:
\begin{widetext}
	\begin{subequations}
		\begin{align}
		\mathcal{G}^\mathrm{R}_\omega(x,y) &= (\mathcal{G}_0^\mathrm{R})_\omega(x,y) + (\mathcal{G}_0^\mathrm{R})_\omega(x,s)\textbf{u}\left(1-(\mathcal{G}_0^\mathrm{R})_\omega(s,s)\textbf{u}\right)^{-1}(\mathcal{G}_0^\mathrm{R})_\omega(s,y),\\
		\nonumber \mathcal{G}^\mathrm{K}_\omega(x,y) &= (\mathcal{G}_0^\mathrm{K})_\omega(x,y) - \mathcal{G}^\mathrm{R}_\omega(x,s)\frac{\mathrm{i}}{2}\bm{\gamma}\mathcal{G}^\mathrm{A}_\omega(s,y) + \mathcal{G}^\mathrm{R}_\omega(x,s)\textbf{u}(\mathcal{G}_0^\mathrm{K})_\omega(s,y)+ (\mathcal{G}_0^\mathrm{K})_\omega(x,s)\textbf{u}^\dagger \mathcal{G}^\mathrm{A}_\omega(s,y)\\
		&~~~~ + \mathcal{G}^\mathrm{R}_\omega(x,s)\textbf{u}(\mathcal{G}_0^\mathrm{K})_\omega(s,s)\textbf{u}^\dagger \mathcal{G}^\mathrm{A}_\omega(s,y),
		\end{align}
		\label{eq:Dyson-quadratic}
	\end{subequations}
\end{widetext}
where the impurity matrices are defined as:
\begin{align}
	\textbf{u} = \left(\begin{matrix}
	u_{L} & 0 \\ 0 & u_{R}
	\end{matrix}\right), \quad \bm{\gamma} = \left(\begin{matrix}
	\gamma_L & 0 \\ 0 & \gamma_R
	\end{matrix}\right).
\end{align}\par
The components of the corresponding scattering matrix $S(k)$ [cf.\ Eq.~\eqref{eq:Scattering-matrix}] are:
\begin{subequations}
	\begin{align}
	t_k &= \frac{v_k^2}{u_Lu_R\ee^{4\mathrm{i}ks}-(u_L-\mathrm{i}v_k)(u_R-\mathrm{i}v_k)},\\
	r^{L,R}_k &= \frac{\ee^{-2\mathrm{i}ks}\left[ u_Lu_R(1-\ee^{4\mathrm{i}ks})-\mathrm{i}v_k(u_{L,R}+\ee^{4\mathrm{i}ks}u_{R,L})\right]}{u_Lu_R\ee^{4\mathrm{i}ks}-(u_L-\mathrm{i}v_k)(u_R-\mathrm{i}v_k)}.
	\end{align}
	\label{eq:Double-barrier-scattering}
\end{subequations}
These can be explicitly read off in the expressions obtained from Eq.~\eqref{eq:Dyson-quadratic}. The resulting impurity Green's functions are given as:
\begin{widetext}
\begin{subequations}
	\begin{align}
		\label{eq:Impurity-Retarded-lj}
		\mathcal{G}^\mathrm{R}_\omega(i,x|j,y) &= \frac{1}{\mathrm{i}v_k}\left[\delta_{ij}\ee^{\mathrm{i}k|x-y|}+S_{ij}\ee^{\mathrm{i}k(x+y)}\right],\\
		\mathcal{G}^\mathrm{K}_\omega(i,x|j,y) &= \frac{1}{\mathrm{i}v_k}\left[ \delta_{ij} h_j \ee^{-\mathrm{i}k(x-y)}+\bar{S}_{ji}h_i\ee^{-\mathrm{i}k(x+y)}+S_{ij}h_j\ee^{\mathrm{i}k(x+y)}+\left(\Delta_{ij}+\sum_l\bar{S}_{jl}S_{il}h_l\right)\ee^{\mathrm{i}k(x-y)}\right],
		\label{eq:Impurity-Keldysh-lj}%\\
	\end{align}
	\label{eq:Impurity-Green}
\end{subequations}
\end{widetext}
where we introduced the Hermitian matrix $\Delta=\mathds{1}-SS^\dagger$ as indicator for nonunitarity of the scattering matrix $S$.
Note that for ideal transmission, i.e. $t_k = 1$ and $r^L_k = r^R_k = 0 = \Delta_{ij}$, the scattering matrix coincides with the Pauli matrix $S_{ij}=\sigma_{ij}^1$ and we obtain the Green's functions of Eqs.~\eqref{eq:Green-unperturbed}. Due to the regularization issues discussed in App.~\ref{sec:Regularization-Green}, the frequency $\omega$ is viewed as positive. Equations~\eqref{eq:Impurity-Green} are thus general expressions for the Green's functions in presence of a lossy impurity with nonunitary scattering matrix $S$.

\begin{figure}[t!]
	\includegraphics[width=0.49\textwidth]{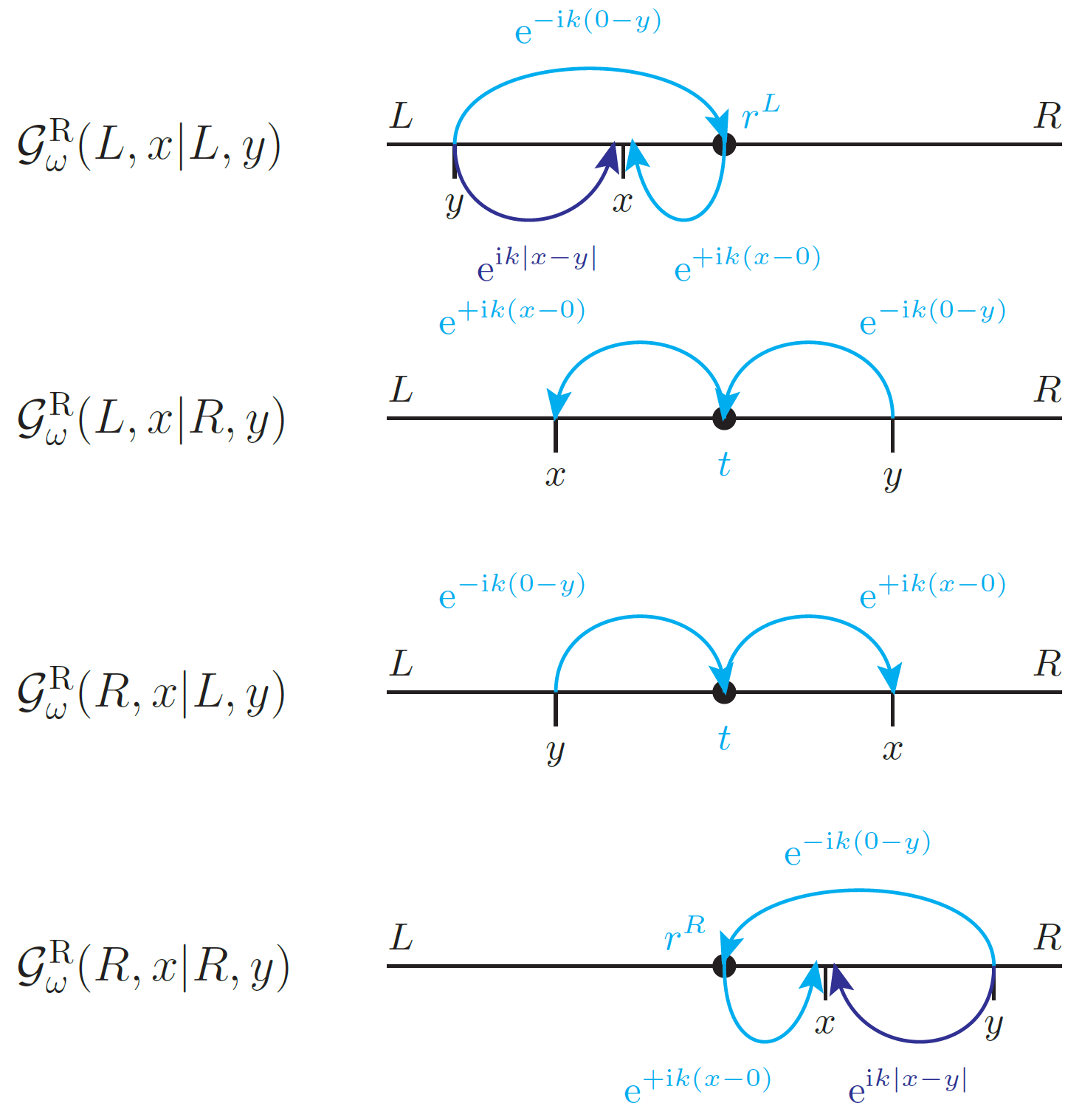}
	\caption{The scattering processes belonging to the retarded Green's function $\mathcal{G}^\mathrm{R}$, Eq.~\eqref{eq:Impurity-Retarded-lj},
		can be understood in a clear physical way: The probability amplitude of a particle traveling from position $y$ to position $x$ inside the same wire $i=j$ is $\ee^{\pm\mathrm{i}k(x-y)}$ where the sign depends on the direction of motion: towards the impurity for $y>x$ and away from the impurity for $x>y$ (dark blue). For processes involving a scattering event at the central impurity from wire $j$ to wire $i$, the scattering amplitude $S_{ij}$ is further taken into account. Thus, there is an overall amplitude $\ee^{+\mathrm{i}k(x-0)}S_{ij}\ee^{-\mathrm{i}k(0-y)} = S_{ij}\ee^{\mathrm{i}k(x+y)}$ (light blue).}
	\label{fig:Retarded-Green-Scheme}
\end{figure}\par
In Fig.~\ref{fig:Retarded-Green-Scheme} we illustrate how the terms of the retarded Green's function can be understood in terms of scattering processes. \par
We finally emphasize that the results illustrated here can be readily extended to a junction of $N>2$ wires by allowing indices $i,j = 1,2,...,N$~\cite{Aristov_2014,Aristov_2017,Aristov_2018,Aristov_2019,nosov2020tunneling}, which represents the building block for networks of quantum wires~\cite{alicea2011non}.

\section{Regularization in first-order perturbation theory\label{sec:Regularization-perturbation}}

In this section, we briefly want to justify why time-local Keldysh Green's functions have to be replaced by the lesser Green's function $\mathcal{G}^\mathrm{K}(t',t')\to 2\mathcal{G}^<(t',t')$ in perturbative Keldysh field theory. To our knowledge, this issue is not discussed sufficiently in the common literature.

In the basis of the fermionic fields $\psi^+$, $\psi^-$ of the plus and minus time contour~\cite{kamenev_2011}, the Keldysh action which corresponds to the Hamiltonian of quartic interactions $H_\mathrm{int} = \int_{x,y}g\psi^\dagger\psi^\dagger\psi\psi$ takes the form:
\begin{align}
	\nonumber \mathrm{i}S_\mathrm{int} &= \!-\frac{\mathrm{i}}{2}\!\int_{x,y,t}\!\! g(x-y)\! \left[ -\bar{\psi}^+(x,t)\bar{\psi}^+(y,t)\psi^+(y,t)\psi^+(x,t)\!\right.\\
	&~~~~~~~~~~~~~~~~\left. +\! \bar{\psi}^-(x,t)\bar{\psi}^-(y,t)\psi^-(y,t)\psi^-(x,t)\right].
	\label{eq:S_int}
\end{align}
Quantum-quantum correlations always vanish, i.e., $\left\langle\psi^q\bar{\psi}^q\right\rangle=0$, and the time-local response functions add up to zero, i.e., $\mathcal{G}^\mathrm{R}(t,t)+\mathcal{G}^\mathrm{A}(t,t)=0$, because of causality. In first-order perturbation theory of the interactions, there are consequently one Hartree and one Fock diagram for each of the response Green's functions $\mathcal{G}^\mathrm{R}$ and $\mathcal{G}^\mathrm{A}$ and two Hartree and Fock terms for the Keldysh Green's function $\mathcal{G}^\mathrm{K}$ (cf.\ Fig.~\ref{fig:First-order-Green}). The first-order corrections of the Green's functions have the form:
\begin{figure}[t!]
	\centering
    \includegraphics[width=0.48\textwidth]{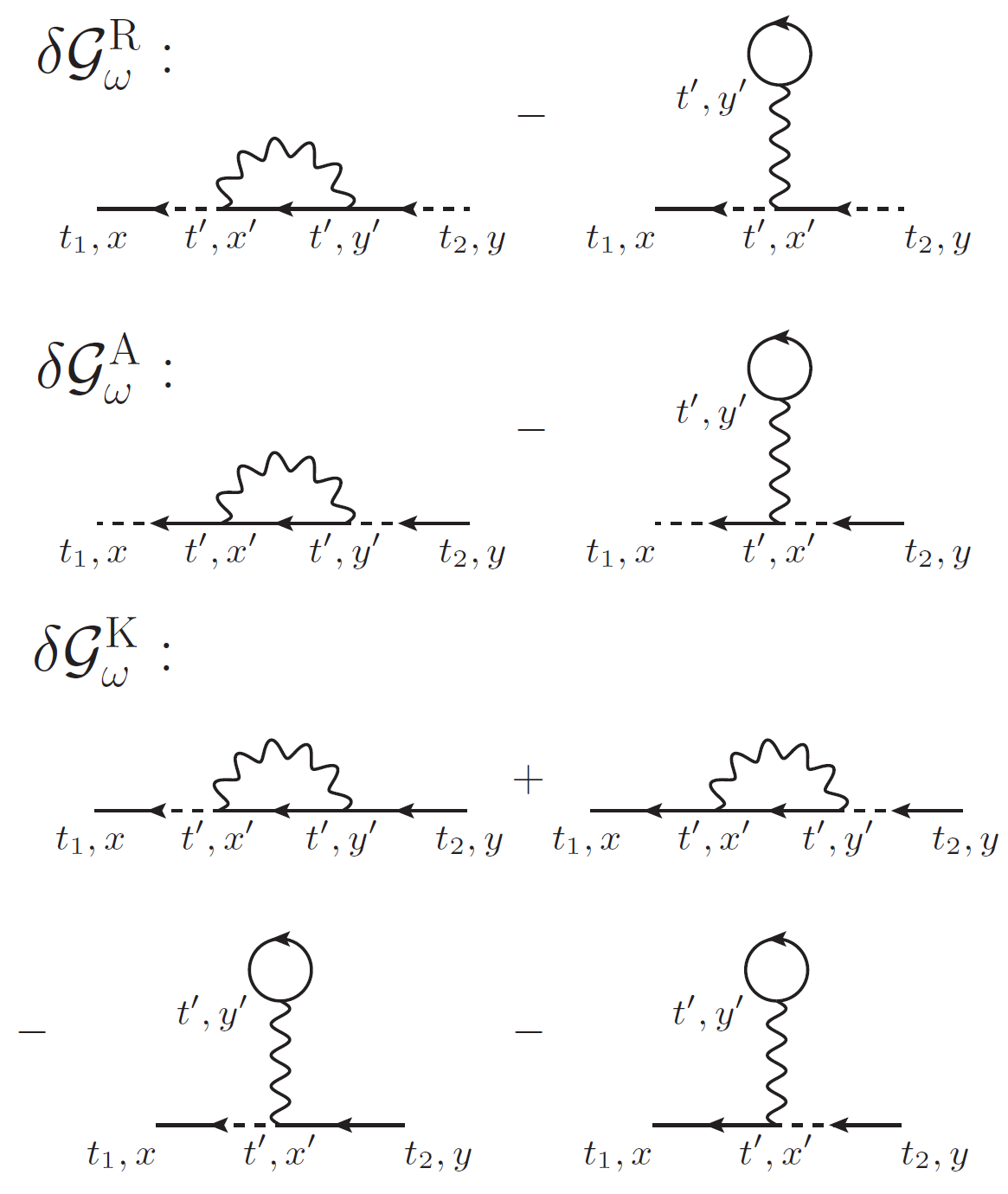}
	\caption{First-order Green's functions in diagrammatic representation.}
	\label{fig:First-order-Green}
\end{figure}
\begin{subequations}
	\begin{align}
	\delta\mathcal{G}^\mathrm{R} &\sim \frac{\mathrm{i}}{2}\int g~ \mathcal{G}_0^\mathrm{R}\mathcal{G}_0^\mathrm{K}\mathcal{G}_0^\mathrm{R},\\
	\delta\mathcal{G}^\mathrm{A} &\sim \frac{\mathrm{i}}{2} \int g~ \mathcal{G}_0^\mathrm{A}\mathcal{G}_0^\mathrm{K}\mathcal{G}_0^\mathrm{A},\\
	\delta \mathcal{G}^\mathrm{K} &\sim \frac{\mathrm{i}}{2}\int g 
	\left[ \mathcal{G}_0^\mathrm{R}\mathcal{G}_0^\mathrm{K}\mathcal{G}_0^\mathrm{K} + 	\mathcal{G}_0^\mathrm{K}\mathcal{G}_0^\mathrm{K}\mathcal{G}_0^\mathrm{A} \right].
	\end{align}
	\label{eq:Keldysh-first-order}
\end{subequations}\par
However, it turns out that the time-local Keldysh Green's function $\mathcal{G}^\mathrm{K}(t',t')$ showing up in the center of the first-order diagrams (cf.\ Fig.~\ref{fig:First-order-Green}) is a pathology of the field theory. Keeping track of the regularization $\delta t$ from the Trotter decomposition in the $+/-$ time contour basis inside the Keldysh action cures this artifact. The more precise interaction term of the Keldysh action $S_\mathrm{int}$, Eq.~\eqref{eq:S_int}, is:
\begin{align}
	\nonumber \mathrm{i}S_\mathrm{int} \!&= \!-\frac{\mathrm{i}}{2}\!\int_{x,y,t}\!\!\!\!\! g(x\!-\! y)\! \left[ -\bar{\psi}^+(x,t\!+\!\delta t)\bar{\psi}^+(y,t\!+\!\delta t)\right.\\
	\nonumber &~~~~~~~~~~~~~~~~~~~~~~~~~~~~~~~\times\psi^+(y,t)\psi^+(x,t)\\
	&\left. \quad + \bar{\psi}^-(x,t)\bar{\psi}^-(y,t)\psi^-(y,t\!+\!\delta t)\psi^-(x,t\!+\!\delta t)\right].
\end{align}
Similar to before, one can write down the first-order correction of the Green's functions $\mathcal{G}^{\sigma\sigma'} = -\mathrm{i}\langle\psi^{\sigma}\bar{\psi}^{\sigma'}\rangle$ in the basis of $+/-$ fields while keeping the time regularization $\delta t$. Only at the end, $\delta t$ is neglected again by the identifications:
\begin{align}
	\nonumber \mathcal{G}^{++}(t',t'+\delta t) &\overset{\delta t \to 0}{\longrightarrow} \mathcal{G}^<(t',t'),\\
	\mathcal{G}^{--}(t'+\delta t,t') &\overset{\delta t \to 0}{\longrightarrow} \mathcal{G}^<(t',t').
\end{align}\par
Finally, one arrives at the corrected expressions for the first-order Green's functions given in the main text [cf.\ Eq.~\eqref{eq:Keldysh-first-order-corrected}].
To conclude, time-local Keldysh Green's functions have to be replaced by the lesser Green's function $\mathcal{G}^\mathrm{K}(t',t')\to 2\mathcal{G}^<(t',t')$ to handle regularization issues in Keldysh field theory properly.\par

\section{Calculation details for deriving the current correction \label{sec:Calculation-Details}}

Here, we provide some details how to derive the explicit dominating part of the first-order current correction $\delta J_{i,\text{sing}}$, given in the main text [cf.\ Eq.~\eqref{eq:J^(1)_div_rewritten}], by starting with Eq.~\eqref{eq:Current-correction-quadratic-start}.

For a structured analysis of $\delta J_i$, Eq.~\eqref{eq:Current-correction-quadratic-start}, we introduce the following notation of the noninteracting Green's functions, Eq.~\eqref{eq:Impurity-Green}:
\begin{align}
	\mathcal{G}^\Xi_\omega(i,x|j,y)&= \frac{1}{\mathrm{i}v_k}\sum_{\sigma_i,\sigma_j}C_{\sigma_i\sigma_j}^\Xi(\omega,x,y) \ee^{\mathrm{i}k(\sigma_ix-\sigma_j y)},
	\label{eq:Green-non-interacting-coefficients}
\end{align}
where the sign factors $\sigma_i,\sigma_j = \pm$ indicate the direction of the annihilated/created fermion in the corresponding scattering process (cf.\ Fig.~\ref{fig:Retarded-Green-Scheme}), and $\Xi = \mathrm{R}, \mathrm{A}, \mathrm{K}$. The coefficients $C_{\sigma_i\sigma_j}^\Xi(\omega,x,y)$ are specific for each Green's function and depend on both the scattering matrix $S_{ij}(k)$ and the Fermi distributions $f_i(k)$ of the Fermi liquid reservoirs. The values for these coefficients $C_{\sigma_i\sigma_j}^\Xi$ are listed in Tab.~\ref{tab:coefficients-Green}.
\begin{table}
	\centering
	\caption{Coefficients $C_{\sigma_i\sigma_j}^\Xi$ inside the four noninteracting Green's functions in Eq.~\eqref{eq:Green-non-interacting-coefficients}. For each sign of momentum direction $\sigma_i, \sigma_j = \pm$ of the annihilated/created fermion and the specific Green's functions $\Xi = \mathrm{R}, \mathrm{A}, \mathrm{K}, <$, there is an implicit sum over repeated indices $l$.} %. \ale{if we remove the lesser, we get it to fit nicely in the column}}
	\begin{tabular}{l|c|c|c|c}
		\hline
		\hline
		$\sigma_i \sigma_j$ & $C_{\sigma_i\sigma_j}^\mathrm{R}$ & $C_{\sigma_i\sigma_j}^\mathrm{A}$ & $C_{\sigma_i\sigma_j}^\mathrm{K}$ & $C_{\sigma_i\sigma_j}^<$ \\
		\hline
		$+-$ & $S_{ij}$ & $0$ & $S_{ij}h_j$ & $-S_{ij}f_j$\\
		$++$ & $\delta_{ij}\theta(x-y)$ & $-\delta_{ij}\theta(y-x)$ & $\Delta_{ij} + \bar{S}_{jl}S_{il}h_l$ & $-\bar{S}_{jl}S_{il}f_l$\\
		$--$ & $\delta_{ij}\theta(y-x)$ & $-\delta_{ij}\theta(x-y)$ & $\delta_{ij}h_j$ & $-\delta_{ij}f_j$\\
		$-+$ & $0$ & $-\bar{S}_{ji}$ & $\bar{S}_{ji}h_i$ & $-\bar{S}_{ji}f_i$\\
		\hline
		\hline
	\end{tabular}
	\label{tab:coefficients-Green}
\end{table}

Due to the definitions of velocities, i.e., $v_k = {\partial \omega}/{\partial k}, v_{k'} = {\partial \omega'}/{\partial k'}$, the frequency integrals over $\omega,\omega'$ inside the current formula $\delta J_i$, Eq.~\eqref{eq:Current-correction-quadratic-start}, can be easily substituted to momentum integrals over $k,k'$ such that the term is rewritten as:
\begin{align}
	\delta J_i = -\frac{1}{4}\sum_l \int_{k,k'}\mathrm{Im}\left[I(x,i,l,k,k')\right],
\end{align}
where the inner term $I(x,i,l,k,k')$ %\marcel{maybe give $I$ explicitly?}
includes the integrals over the inner positions $x', y'$ and the different products over three Green's functions $(\partial_x \mathcal{G}^{\Xi_1})\mathcal{G}^{\Xi_2}\mathcal{G}^{\Xi_3}$. Each of the three integrated Green's functions $(\partial_x \mathcal{G}^{\Xi_1})\mathcal{G}^{\Xi_2}\mathcal{G}^{\Xi_3}$ contains two sign factors $\sigma_i, \sigma_j$ such that in total $2^6 = 64$ terms arise for each product of Green's functions due to the six sign factors $\sigma_1, ..., \sigma_6 = \pm 1$. We are interested in the current corrections at the end of the wire, i.e., outside of the interacting region $\delta J_i(x>L/2)$. The causal structure of the Green's functions (cf.\ Tab.~\ref{tab:coefficients-Green}) yields to vanishing contributions for half the terms as $C^\mathrm{R}_{-+}=0=C^\mathrm{R}_{--}$ and $C^\mathrm{A}_{+-}=0=C^\mathrm{A}_{--}$ for $x>L/2>x',y'$. Further terms vanish as they sum up to a real number before taking the imaginary part $\mathrm{Im}\left[(\partial_x\mathcal{G}^{\Xi_1})\mathcal{G}^{\Xi_2}\mathcal{G}^{\Xi_3}\right]$. Finally, only the $8$ terms with $\sigma_1 = \sigma_6 = +1$ survive, which is a quarter of the before-mentioned $64$ terms. Physically this corresponds to the fact that outside of the interacting region $x>L/2$ the incoming part of the current is not affected by the interactions inside the wires such that only the outgoing part of the current is renormalized.\par

The integrals over the inner position variables $x',y'$ are executed by the following three steps:\par

1. We transform the position variables $x',y'$ into a center-of-mass $X=\frac{1}{2}(x'+y')$ and a relative coordinate $\xi = x'-y'$ such that we get an integral expression of the form:
\begin{align}
	\int_s^{L/2}\mathrm{d}X\int_{-l(X)}^{l(X)}\mathrm{d}\xi ~g_l(\xi)\ee^{\mathrm{i}(K\xi+2\tilde{K}X)},
\end{align}
where the boundaries $\pm l(X)$ depend on the center-of-mass coordinate $X$ and $K=K(k,k')$, $\tilde{K}=\tilde{K}(k,k')$ are linear combinations of the momenta $k,k'$.\par
2. As the interaction potential $g_l(\xi)$ is short-ranged, we can expand the limits $l(X)\to \infty$ and solve the integral over the relative coordinate $\xi$ by making use of the Fourier transform $\tilde{g}_l(K)=\int_\xi g_l(\xi)\ee^{-\mathrm{i}K\xi}$.\par
3. The integral over the center-of-mass coordinate $X$ can be approximated as follows:
\begin{align}
	\int_s^{L/2}\mathrm{d}X~\ee^{2\mathrm{i}\tilde{K}X}= \frac{\ee^{\mathrm{i}\tilde{K}L}-\ee^{2\mathrm{i}\tilde{K}s}}{2\mathrm{i}\tilde{K}}\overset{\tilde{K}L\to\infty,\tilde{K}s\to 0}{\longrightarrow}-\frac{1}{\tilde{K}+\mathrm{i}0^+}.
\end{align}
The regularization $0^+$ is caused by physical infrared cutoffs like a finite length $L$ or a finite temperature $T$. %In the main text we refer that scale to the inverse length of the wire, i.e., $0^+\to 1/L$.\par
An overview of the resulting values of $K$ and $\tilde{K}$ for the different values of the sign factors $\sigma_2, ..., \sigma_5$ [cf.\ Eq.~\eqref{eq:Green-non-interacting-coefficients}] in the products of Green's functions $(\partial_x\mathcal{G}^{\Xi_1})\mathcal{G}^{\Xi_2}\mathcal{G}^{\Xi_3}$ inside the Eq.~\eqref{eq:Current-correction-quadratic-start} for the correction of the current $\delta J_i$ is provided in Tab.~\ref{tab:Hartree-Fock-exponent-V}.
\begin{table}
	\centering
	\caption{Momentum terms $\tilde{K}$ and Hartree and Fock potentials corresponding to different inner sign factors $\sigma_2, ..., \sigma_5$.}
	\begin{tabular}{l|l|l|l}
		\hline
		\hline
		$\sigma_2\sigma_3\sigma_4\sigma_5$ & $\tilde{K}$ & $\tilde{g}(K)$ Fock & $\tilde{g}(K)$ Hartree \\
		\hline
		$-+-+$ & $k+k'$ & $\tilde{g}(0)$ & $\tilde{g}(k-k')$ \\
		$-+--$ & $k'$ & $\tilde{g}(k)$ & $\tilde{g}(-k')$ \\
		$-+++$ & $k$  & $\tilde{g}(k')$ & $\tilde{g}(k)$ \\
		$-++-$ & $0$   & $\tilde{g}(k+k')$ & $\tilde{g}(0)$ \\
		$---+$ & $k$  & $\tilde{g}(-k')$ & $\tilde{g}(k)$ \\
		$----$ & $0$   & $\tilde{g}(k-k')$ & $\tilde{g}(0)$ \\
		$--++$ & $k-k'$ & $\tilde{g}(0)$ & $\tilde{g}(k+k')$ \\
		$--+-$ & $-k'$ & $\tilde{g}(k)$ & $\tilde{g}(k')$ \\
		$++-+$ & $k'$ & $\tilde{g}(-k)$ & $\tilde{g}(-k')$ \\
		$++--$ & $-k+k'$ & $\tilde{g}(0)$ & $\tilde{g}(-k-k')$ \\
		$++++$ & $0$   & $\tilde{g}(-k+k')$ & $\tilde{g}(0)$ \\
		$+++-$ & $-k$ & $\tilde{g}(k')$ & $\tilde{g}(-k)$ \\
		$+--+$ & $0$   & $\tilde{g}(-k-k')$ & $\tilde{g}(0)$ \\
		$+---$ & $-k$ & $\tilde{g}(-k')$ & $\tilde{g}(-k)$ \\
		$+-++$ & $-k'$ & $\tilde{g}(-k)$ & $\tilde{g}(k')$ \\
		$+-+-$ & $-k-k'$ & $\tilde{g}(0)$ & $\tilde{g}(-k+k')$ \\
		\hline
		\hline
	\end{tabular}%
	\label{tab:Hartree-Fock-exponent-V}
\end{table}\par
The eight nonvanishing terms correspond to the following combinations of sign factors:
\begin{align}
    \nonumber &(\sigma_2,\sigma_3,\sigma_4,\sigma_5) =\\
    \nonumber & (+,+,+,-), (+,+,-,-), (+,-,-,-), (+,-,+,-),\\
    &(-,-,-,+), (-,-,+,+), (-,+,+,+), (-,+,-,+).
\end{align}
By using the definition of the Hermitian loss matrix $\Delta_{li}$ and the distribution function $h_j = 1-2f_j$ we have $\Delta_{li}+\sum_j\bar{S}_{ij}S_{lj}h_j-\delta_{li}h_l = -\sum_j \left(\bar{S}_{ij}S_{lj}-\delta_{ij}\delta_{lj}\right)2f_j$. With further manipulations according to $\mathrm{Im}(z)=-\mathrm{Im}(\bar{z})$ and $\tilde{g}(k)=\tilde{g}(-k)$, the terms $(--++)$ and $(++--)$ can be combined to $\delta J_{i,\text{sing}}$ \eqref{eq:J_div^(1)}, $(-+-+)$ and $(+-+-)$ to $\delta J_{i,\mathrm{reg},1}$, Eq.~\eqref{eq:J_back^(1)}, and finally $(-+++)$ $(---+)$, $(+++-)$ and $(+---)$ to $\delta J_{i,\mathrm{reg},2}$, Eq.~\eqref{eq:J_chem^(1)}. So all the nonvanishing current contributions are summarized in the following three contributions:
\begin{widetext}
	\begin{subequations}
		\begin{align}
		\delta J_{i,\text{sing}}&= -\sum_{j,l}\int_{k,k'}\left( \tilde{g}_l(0)\!-\! \tilde{g}_l(k+k') \right)\mathrm{Re}\left[S_{il}\bar{S}_{ll}'(\bar{S}_{ij}S_{lj}\!-\! \delta_{ij}\delta_{il})\frac{f_l'f_j}{k-k'+\mathrm{i}0^+}\right],
		\label{eq:J_div^(1)}\\
		\delta J_{i,\mathrm{reg},1}&= -\sum_{j,l}\int_{k,k'}\left( \tilde{g}_l(0) \!-\! \tilde{g}_l(k\!-\! k')\right)\mathrm{Re}\left[S_{il}S_{ll}'(\bar{S}_{ij}S_{lj}\!-\! \delta_{ij}\delta_{il})\frac{f_l'f_j}{k+k'+\mathrm{i}0^+} \right],
		\label{eq:J_back^(1)}\\
		\delta J_{i,\mathrm{reg},2}&= -\sum_{j,l,n}\int_{k,k'}\left( \tilde{g}_l(k')\!-\! \tilde{g}_l(k) \right)\left(|S_{ln}'|^2+\delta_{ln}\right)\mathrm{Re}\left[S_{il} (\bar{S}_{ij}S_{lj}\!-\! \delta_{ij}\delta_{il})\frac{f_n'f_j}{k+\mathrm{i}0^+} \right].
		\label{eq:J_chem^(1)}
		\end{align}
		\label{eq:Current-Glazman-all}
	\end{subequations}
\end{widetext}
For the sake of clarity in notation, we omitted the momentum dependence of the scattering matrix and the population functions, i.e., $S_{ij}=S_{ij}(k)$, $S'_{ij}=S_{ij}(k')$ and $f_i=f_i(k)$, $f'_i=f_i(k')$.

In the first term, Eq.~\eqref{eq:J_div^(1)}, the contribution proportional to $\delta_{ij}\delta_{il}$ vanishes which can be seen from interchanging the momentum variables $k\leftrightarrow k'$. Consequently, we have the expression given in the main text [cf.\ Eq.~\eqref{eq:J^(1)_div_rewritten}].

We are interested in the limit where the voltage is the smallest energy scale and serves as a physical infrared cutoff. We assume $T=0$ so that $f_l(k) = \theta(k_l-k)$, and infinite system size $L \to +\infty$.

The universal behavior of the integral expression for the current $\delta J_{i,\text{sing}}$ can be discussed in the limit of small voltages $V = \mu_L-\mu_R\ll\mu$ between the two reservoirs, i.e., similar Fermi momenta on the two sites of the impurity $k_L\simeq k_R$. 

We assume that the momentum dependence of the interaction potential $\tilde{g}_l(k+k')$ and the scattering amplitudes $S_{jl}(k)$ is smooth near the average Fermi momentum $k_\mathrm{F} =\left(k_L+k_R\right)/2$ so that their values can be well approximated at the Fermi momentum as the integral is dominating there. The double momentum integral can then be approximated by the following form:
\begin{align}
\int_0^{k_j}\!\!\!\mathrm{d}k\!\int_0^{k_l}\!\!\!\mathrm{d}k'\frac{1}{k\!-\!k'\!+\!\mathrm{i}0^+}\simeq -\ii\pi k_\mathrm{F} + \left(k_j\! -\! k_l\right)\ln\left\vert\frac{k_\mathrm{F}}{k_j\!-\!k_l}\right\vert.
\label{eq:double-k-integral}
\end{align}
By extracting only the logarithmic part, we obtain the current correction $\delta J_{i,\mathrm{sing}}$ given in the main text [cf.\ Eq.~\eqref{eq:Current-correction-simplification}].

The derivation of the correction for scattering amplitudes $\delta S_{ij}$, Eq.~\eqref{eq:Correction-S_jl}, from the correction of the retarded Green's function $\delta G^R$, Eq.~\eqref{eq:GR^(1)}, works in an analogous way to the derivation of the current $\delta J_{i,\mathrm{sing}}$, Eq.~\eqref{eq:J^(1)_div_rewritten}, from the Keldysh Green's function $\delta G^K$ and has been extensively discussed in the literature~\cite{matveev1993tunneling,Yue1994,Froeml2019,Froeml2020,muller2021shape}. Similar to the integral expression \eqref{eq:double-k-integral}, the logarithmic scaling for the scattering amplitude, Eq.~\eqref{eq:Correction-S_jl-compactified}, can be extracted by the following approximation:
\begin{align}
    \int_0^{k_j}\dd k' \frac{S(k')}{k-k'+\ii 0^+}\simeq S(k)\ln\left\vert\frac{k}{k-k_j}\right\vert.
    \label{eq:single-k-integral}
\end{align}

\section{Conductance RG equation}

The exact solution of the conductance RG equation \eqref{eq:RG-equations} for the noninteracting conductance value $G_0 = |t|^2+\eta/2$, which is valid for a $\delta$-shaped impurity, yields:
\begin{align}
    \nonumber G(\Lambda) &= \frac{\sqrt{1-2\eta}}{2}\tanh\left[\mathrm{artanh}\frac{2G_0-1}{\sqrt{1-2\eta}}-\alpha\Lambda\sqrt{1-2\eta}\right]\\
    &\quad+\frac{1}{2}.
    \label{eq:G(Lambda)_exact}
\end{align}
From the condition $t=1+r$ of a $\delta$ impurity [cf.\ Eq.~\eqref{eq:Scattering-delta}], we have $\eta = -2\mathrm{Re}(\bar{r}t)$ and thus:
\begin{align}
    \nonumber \eta^2 &= 4\mathrm{Re}(\bar{r}t)^2 = 4|rt|^2\cos(2(\phi_r-\phi_t))\\
    &\leq 4|rt|^2 = 4|t|^2\left(1-|t|^2-\eta\right),
\end{align}
where $\phi_r$ and $\phi_t$ are the phases of the scattering amplitudes $r$ and $t$, respectively. Inserting the noninteracting conductance $G_0 = |t|^2+\eta/2$ yields:
\begin{align}
    G_0\left(1-G_0\right)\geq\frac{\eta}{2},
    \label{eq:Inequality_G}
\end{align}
which allows only initial values $G_0$ inside the parabola given by the fixed points $G^*$. Thus, the inequality \eqref{eq:Inequality_G} is valid for the renormalized conductance $G(\Lambda)$, too.

\section{Renormalization of scattering amplitudes \label{sec:RG-scattering}}

Let us write down the first-order corrections for the scattering amplitudes $\delta S_{ij}(k)$, Eq.~\eqref{eq:Correction-S_jl}, of a single wire with a dissipative impurity and $2\times 2$ scattering matrix $S(k)$, Eq.~\eqref{eq:Scattering-matrix}. For zero voltage, there is only one distinct Fermi momentum $k_\mathrm{F}$ and the first-order corrections of the scattering amplitudes are:
\begin{subequations}
	\begin{align}
	\delta t_k &= - \frac{\alpha}{2} \left( r^L_k \bar{r}^L t_k + t_k \bar{r}^R\rho^R_k\right) \ln \left\vert\frac{k_\mathrm{F}}{k-k_\mathrm{F}}\right\vert,\\
	\delta r^{L,R}_k &= - \frac{\alpha}{2} \left(r^{L,R}_k\bar{r}^{L,R}r^{L,R}_k+t_k\bar{r}^{R,L}t_k-r^{L,R}\right)\ln \left\vert\frac{k_\mathrm{F}}{k-k_\mathrm{F}}\right\vert.
	\end{align}
\end{subequations}
These terms can be nicely understood by the occurring scattering processes visualized in Fig.~\ref{fig:First-order-current}.

At the Fermi momentum $k_\mathrm{F}$, i.e., $t\equiv t_{k_\mathrm{F}}$ and $r^{L,R}\equiv r^{L,R}_{k_\mathrm{F}}$, the following RG equations are derived by setting the RG scale parameter $\Lambda = \ln\vert{k_\mathrm{F}}/({k-k_\mathrm{F}})\vert$:
\begin{subequations}
	\begin{align}
	\frac{\mathrm{d}t}{\mathrm{d}\Lambda} &= - \frac{\alpha}{2} \left(|r^L|^2+|r^R|^2\right)t,\\
	\frac{\mathrm{d}r^{L,R}}{\mathrm{d}\Lambda} &= - \frac{\alpha}{2}\left(|r^{L,R}|^2r^{L,R}+t^2\bar{r}^{R,L}-r^{L,R}\right).
	\end{align}
\end{subequations}
For a symmetric impurity $r^L=r^R$, one obtains the RG flow equations for the transmission $|t|^2$ and loss probability $\eta = 1- |r|^2-|t|^2$ of a dissipative $\delta$ impurity:
\begin{subequations}
	\begin{align}
	\frac{\mathrm{d}|t|^2}{\mathrm{d}\Lambda} &= -2\alpha|t|^2\left(1-|t|^2-\eta\right),\\
	\frac{\mathrm{d}\eta}{\mathrm{d}\Lambda} &= \alpha \left(|t|^2+\frac{3}{2}\eta-1\right)\eta .
	\end{align}
	\label{eq:RG-t-eta}
\end{subequations}
These equations were broadly analyzed in our previous work~\cite{Froeml2019,Froeml2020}. There it was shown that the renormalized loss probability vanishes at the Fermi momentum $\eta = 0$ for both kinds of interactions. In particular, for repulsive interactions $\alpha>0$ the vanishing of the loss probability is understood by the fluctuation-induced quantum Zeno effect.

Now let us write down the RG flow of the combination of the zeroth-order conductance, namely $|t|^2+\eta/2$:
\begin{align}
    \nonumber\frac{\dd}{\dd\Lambda} \left(|t|^2+\frac{\eta}{2}\right) &= -2\alpha\left[\frac{\eta}{2}-\left(|t|^2+\frac{\eta}{2}\right)\left(1-|t|^2-\frac{\eta}{2}\right)\right]\\
    &\quad -\alpha\left[\frac{\eta}{2}(|t|^2-1)+\frac{11}{4}\eta^2\right].
    \label{eq:RG_T+eta/2}
\end{align}
This RG equation is valid for the scattering probabilities at the Fermi momentum, where the scale parameter $\Lambda = \ln|k_\mathrm{F}/(k-k_\mathrm{F})|$ was introduced after the voltage was set to zero. Eq.~\eqref{eq:RG_T+eta/2} is not equivalent to the RG equation \eqref{eq:RG-equations} of the conductance $G$ as that was derived from the current and in terms of a scale parameter $\Lambda = \ln|V/\mu|$. To describe dissipative transport properties properly, one cannot exclusively rely on the scattering probabilities renormalized at the Fermi energy as the loss current is also governed by states far below that. 

\section{Fluctuation-induced QZE in a Y-junction \label{sec:Y-junction}}

This section mainly summarizes and comments the discussion given in Refs.~\onlinecite{Aristov_2014,Aristov_2017,Aristov_2018,Aristov_2019}. The Y-junction, consisting of three wires $i = 1,2,3$ coupled to each other at a central impurity, is characterized by the following unitary scattering matrix:
\begin{align}
S(k)=\left(\begin{matrix}
r_k & t_k & \tau_k \\ t_k & r_k & \tau_k \\ \tau_k & \tau_k & \rho_k
\end{matrix}\right).
\end{align}
The transmission probability $|\tau_k|^2$ plays the role of the loss probability $\eta_k$ of a dissipative wire and fulfills the same conditions, i.e., $|\tau_k|^2 = 1-|t_k|^2-|r_k|^2 = -2\mathrm{Re}(\bar{r}_kt_k)=\eta_k$.\par
In the case of a noninteracting third wire, i.e., $\alpha_3 = 0$, from Eq.~\eqref{eq:Current-correction-simplification} the logarithmically scaling parts of the three currents pointing toward the reservoirs are given by:
\begin{widetext}
\begin{subequations}
	\begin{align}
		\delta J_{1,\text{sing}} &= \alpha \left(2|rt|^2-\frac{1}{2}|\tau|^4\right)V_a\ln\left\vert\frac{2\mu}{V_a}\right\vert + \alpha|r\tau|^2V_{13}\ln\left\vert\frac{2\mu}{V_{13}}\right\vert -\frac{\alpha}{2}|\tau|^4V_{23}\ln\left\vert\frac{2\mu}{V_{23}}\right\vert,\\
		\delta J_{2,\text{sing}} &= \alpha \left(-2|rt|^2+\frac{1}{2}|\tau|^4\right)V_a\ln\left\vert\frac{2\mu}{V_a}\right\vert-\frac{\alpha}{2}|\tau|^4V_{13}\ln\left\vert\frac{2\mu}{V_{13}}\right\vert+  \alpha|r\tau|^2V_{23}\ln\left\vert\frac{2\mu}{V_{23}}\right\vert,\\
		\delta J_{3,\text{sing}} &= \alpha |\tau|^2\left(\frac{1}{2}|\tau|^2-|r|^2\right)\left(V_{13}\ln\left\vert\frac{2\mu}{V_{13}}\right\vert+V_{23}\ln\left\vert\frac{2\mu}{V_{23}}\right\vert\right).
	\end{align}
	\label{eq:Y-junction-J1-J2-J3}
\end{subequations}
\end{widetext}
With the translation of voltages $V_{13}=V_b + \frac{1}{2}V_a$, $V_{23} = V_b + \frac{1}{2}V_a$, the definition of the logarithmic scales
\begin{align}
	\Lambda_a = \ln\left\vert\frac{2\mu}{V_a}\right\vert, \Lambda_{b+} = \ln\left\vert\frac{2\mu}{V_b+\frac{1}{2}V_a}\right\vert, \Lambda_{b-} = \ln\left\vert\frac{2\mu}{V_b-\frac{1}{2}V_a}\right\vert,
\end{align}
and the structure of the currents $\frac{1}{2}\left(J_2-J_1\right) = G_a V_a + G_{ab}V_b$, $J_3 = G_{ba}V_a + G_b$, the corrections for the conductances can be 
obtained as specific coefficients in %\eqref{eq:Y-junction-J-main-J-sec}
Eq.~\eqref{eq:Y-junction-J1-J2-J3}. With the noninteracting conductances $G_{0,a} = |t|^2+|\tau|^2/2$, $G_{0,b}=2\eta$ and $G_{0,ab}=0=G_{0,ba}$, the corrections for the conductances can be brought into self-consistent expressions~\cite{Aristov_2014,Aristov_2017}:
\begin{subequations}
	\begin{align}
		\nonumber \delta G_a &= 
		2\alpha\left[\frac{G_{0,b}}{4}-G_{0,a}(1-G_{0,a})\right]\Lambda_a \\
		&\quad- \alpha\frac{G_{0,b}}{8}\left(1-G_{0,a}\right)\left(\Lambda_{b-}+\Lambda_{b+}\right) ,\\
		\delta G_b &= %\alpha |\tau|^2\left(\frac{1}{2}|\tau|^2-|r|^2\right)\left(\Lambda_{b+}+\Lambda_{b-}\right)=
		\alpha\frac{G_{0,b}}{2}\left[\frac{G_{0,b}}{2}-(1-G_{0,a})\right]\left(\Lambda_{b+}+\Lambda_{b-}\right),\\
		\delta G_{ab} &=
		-\alpha\frac{G_{0,b}}{4}\left(1-G_{0,a}\right)\left[\Lambda_{b-}-\Lambda_{b+}\right],\\
		\delta G_{ba} &= %\alpha |\tau|^2\left(\frac{1}{2}|\tau|^2-|r|^2\right)\left(\Lambda_{b+}-\Lambda_{b-}\right)=
		\alpha\frac{G_{0,b}}{2}\left[\frac{G_{0,b}}{2}-(1-G_{0,a})\right]\left(\Lambda_{b+}-\Lambda_{b-}\right) .
	\end{align}
\end{subequations}
Note that in the limit of a large bias voltage from the main wire to the third reservoir $V_b \to \mu \Rightarrow \Lambda_{b\pm} \to 0$, the conductance $G_b$ along the secondary wire does not renormalize and we are left with the conductance $\delta G_a$ of the main wire which can be identified with the conductance $\delta G$ in the main text [cf.\ Eq.~\eqref{eq:Ga-Gb-dissipative}] as $G_b$ corresponds to $2\eta$ in that case.\par
However, in the limit $V_a,V_b\to 0$ where the logarithmic scales can be roughly identified with each other, i.e., $\Lambda \equiv \Lambda_a \simeq \Lambda_{b+}\simeq \Lambda_{b-}$, we receive the RG flow equations extensively discussed by Aristov and Wölfle~\cite{Aristov_2017}:
\begin{subequations}
	\begin{align}
		\frac{\mathrm{d}G_a}{\mathrm{d}\Lambda} &= 2\alpha\left(\frac{G_b}{4}-G_a(1-G_a)\right)-\alpha\frac{G_b}{4}(1-G_a),\\
		\frac{\mathrm{d}G_b}{\mathrm{d}\Lambda} &= \alpha~{G_b}\left(\frac{G_b}{2}-(1-G_a)\right),\\
		\frac{\mathrm{d}G_{ab}}{\mathrm{d}\Lambda} &= 0 = \frac{\mathrm{d}G_{ba}}{\mathrm{d}\Lambda}.
	\end{align}
	\label{eq:Y-junction-RG}
\end{subequations}\par
In that case, the corresponding RG flow recovers the fixed points $(G_a^*,G_b^*)=(0,0),(1,0)$ of ideal reflectivity and transparency from the Kane-Fisher problem~\cite{kane1992transmission,kane1992transport}. Our new insight is that the RG equations of conductances in a Y-junction with small voltages $V_a, V_b$, Eq.~\eqref{eq:Y-junction-RG} are equivalent to the RG equation of a dissipative impurity at zero voltage, Eq.~\eqref{eq:RG-t-eta}. As long as the voltage $V_b$ between the main wire and the third reservoir is kept small, indeed, we can identify the renormalized values of the Y-junction conductances with the scattering probabilities:
\begin{subequations}
	\begin{align}
		\frac{\mathrm{d}G_a}{\mathrm{d}\Lambda} &= \frac{\mathrm{d}}{\mathrm{d}\Lambda}\left(|t|^2+\frac{1}{2}\eta\right),\\
		\frac{\mathrm{d}G_b}{\mathrm{d}\Lambda} &= \frac{\mathrm{d}}{\mathrm{d}\Lambda}2\eta.
	\end{align}
	\label{eq:Identification}
\end{subequations}
Consequently, the fluctuation-induced quantum Zeno effect of a dissipative impurity is also found in a Y-junction with small voltage biases.\par
The identification of conductances and scattering probabilities, Eq.~\eqref{eq:Identification}, is broken in a truly dissipative wire as discussed in the main text.

%\bibliographystyle{apsrev4-2}
%\bibliography{Literature-Master-Thesis}
\bibliography{main}

\end{document}